\documentclass[twocolumn,prb,floatfix]{revtex4}
\usepackage{epsfig}
\usepackage{epsf}
\usepackage{color}
\usepackage{hyperref}

\newcommand{\rr}{{\bf r}}
\newcommand{\rrp}{{\bf r}^\prime}
\newcommand{\rrpp}{{\bf r}^{\prime \prime}}
\newcommand{\dd}{{\rm d}}
\newcommand{\oogut}{{\"{O}\u{g}\"{u}t }}
\newcommand{\bra}{\langle}
\newcommand{\ket}{\rangle}
\newcommand{\vcoul}{V}

\begin{document}

\title{Optical excitations in
  organic molecules, clusters and defects studied by first-principles
  Green's function methods}

\author{Murilo L. Tiago$^{(a)}$ and James R. Chelikowsky$^{(a,b)}$} 
\affiliation{
  $^{(a)}$ Center for Computational Materials, Institute
  for Computational Engineering and Sciences, University of Texas,
  Austin, TX 78712, USA.\\
 $^{(b)}$ Departments of Physics and Chemical Engineering,
  University of Texas, Austin, TX 78712, USA.}

\date{\today}

\begin{abstract}

Spectroscopic and optical properties of nanosystems and point defects
are discussed within the framework of Green's function methods.
We use an approach based on evaluating the self-energy in the
so-called GW approximation and solving the Bethe-Salpeter equation in
the space of single-particle transitions. Plasmon-pole models or
numerical energy integration, which have been used in most of the
previous GW calculations, are not used. Fourier transforms of the
dielectric function are also avoided. This approach is applied to
benzene, naphthalene, passivated silicon clusters (containing more
than one hundred
atoms), and the F center in LiCl. In the latter, excitonic effects and
the $1s \rightarrow 2p$ defect line are identified in the
energy-resolved dielectric function. We also compare optical spectra
obtained by solving the Bethe-Salpeter equation and by using
time-dependent density
functional theory in the local, adiabatic approximation. From
this comparison, we conclude that both methods give similar
predictions for optical excitations in benzene and naphthalene, but
they differ in the spectra of small silicon clusters. As cluster size
increases, both methods predict very low cross section for
photoabsorption in the optical and near ultra-violet ranges. For the
larger clusters, the computed cross section shows a slow increase as
function of photon frequency.
Ionization potentials and electron affinities of molecules and
clusters are also calculated.
\end{abstract}

\maketitle

%%%%%%%%%%%%%%%%%%%%%%%%%%%%%%%%%%%%%%%%%%%%%%%%%%%%%%%%%%%%%%%%%%%%%%%%%%
%%%%%%%%%%%%%%%%%%%%%%%%%%%%%%%%%%%%%%%%%%%%%%%%%%%%%%%%%%%%%%%%%%%%%%%%%%
\newpage

\section{Introduction}\label{introduction}

Universality and predictive power are characteristic features of good
first-principles theories. Such theories are invaluable as guides for
experimental work, especially when knowledge about the system being
investigated is limited. Density-functional theory (DFT) is recognized as state-of-the-art theory for the
calculation of various ground-state properties of electronic systems
from first principles, having been applied to a variety of different
systems ranging from crystalline semiconductors to surfaces and
nanostructures \cite{martin}. In contrast, DFT has known
difficulties in predicting quantities associated with excited states,
which is not surprising since it was first proposed as a ground-state
theory. As an example, the electronic band structure provided by DFT
consistently shows an underestimated gap between valence and
conduction bands \cite{hybertsen86,hedin69,aulbur00,martin}. As
consequence, the electronic band structure provided by DFT gives poor
predictions for the onset of photoemission and inverse
photoemission \cite{hybertsen86,aulbur00}.
Exact-exchange functionals improve results considerably for molecular
systems \cite{gorling99}, but more work is needed in crystalline
systems \cite{rinke05}. Gap narrowing also affects the linear
optical spectrum, when calculated from single-electron transitions
from the valence band to the conduction band \cite{onida02,aulbur00}.

Spectroscopic and optical properties of semiconductors have been
successfully predicted from {\it ab initio} Green's function methods
\cite{aulbur00,onida02}. The GW approximation has been used to provide
the quasiparticle band structure of semiconductors, large-gap
insulators, metals, surfaces, nanostructures and other materials
\cite{aulbur00}. By solving the Bethe-Salpeter equation for electrons
and holes, the linear optical spectrum can also be obtained, providing 
reliable results for the optical gap and excitonic effects
\cite{rohlfing00,onida02,rohlfing00_1,albrecht98,tiago03}. Since knowledge of the electron self-energy
is required in solving the Bethe-Salpeter equation (BSE), this method is often refered to as
GW+BSE \cite{grossman01}. Although powerful, the GW+BSE method is
very complex when applied to non-periodic systems, especially when discrete
Fourier analysis is not applicable and approximations such
as the generalized plasmon pole model \cite{hybertsen86} do not lead
to substantial simplification. Some materials that fall in this
category are confined systems: quantum dots, clusters,
molecules, nanostructures. On the other hand, nanostructures are now the
subject of intense work due to promising technological applications
and recent advances  in the preparation of artificial
nanostructures \cite{nano,soloviev01}. At the same time, significant effort has been
dedicated on the theory side. Some of the difficulties one may
encounter by applying the GW approximation to large molecules were
emphasized recently \cite{niehaus05,hahn05}. So far, reliable calculations of
electronic self-energy within the GW approximation have been done for
confined systems with up to a few tens of atoms
\cite{rohlfing00,rohlfing00_1,ismail-beigi03,benedict03,ishii01}. Similarly, the
Bethe-Salpeter equation has been solved for a limited number of
confined systems. During the last decade, optical properties of confined
systems have been investigated within time-dependent DFT, often with
very good results \cite{vasiliev02,onida02}. Time-dependent density
DFT in the local, adiabatic approximation (TDLDA) is formally simpler
than the GW+BSE theory, but at the same time directed towards solving the
same problem: predicting optical properties and neutral excitations in
an electronic system. A drawback of TDLDA is that it reportedly fails to predict
correct optical gaps and excitonic effects in extended systems
\cite{onida02,reining02}.
Alternative TDDFT functionals which include many-body effects have been
proposed \cite{reining02,delsole05,marini03,bruneval05,sottile05}.

As mathematical simplification, applications of the GW+BSE method in
nanostructures frequently impose artificial periodicity by placing the
electronic system in a large supercell \cite{ismail-beigi03}. This is a perfectly
justifiable procedure, but it has the obvious deficiency of increasing
the computational effort when supercells are required to be large. A
more efficient procedure would be to take advantage of properties
intrinsic to confined systems and design a numerical implementation of
such Green's function methods, and we now propose one 
such implementation. A key ingredient is electronic
screening. We incorporate it into the theory by computing the electron
polarizability function in two approximations: random-phase
approximation (RPA) and TDLDA. The electron self-energy is calculated
from first principles and the Bethe-Salpeter equation is set and
solved in the space of single-particle transitions. In this framework,
many-body functions are expressed as matrices in this transition
space and in energy (frequency) representation. Fourier transforms are
not needed, and energy integrations are evaluated by integrations over
poles. Portions of this work have been reported earlier \cite{tiago05}.

The methodology proposed here is applied to a number of interesting
cases. One of them is isolated oligoacenes, for which extensive experimental
data is available but the only GW-based analysis known to us
have been published recently \cite{niehaus05}. The second application is
in silicon nanoclusters, studied previously within TDLDA
\cite{vasiliev02} and, for the smaller clusters, GW+BSE
\cite{rohlfing00,benedict03}. Here, we investigate clusters with
bulk-like crystalline structure and report benchmark GW+BSE
calculations for clusters with more than one hundred silicon atoms. 
The last application is in the F-center defect in
LiCl. This is a challenging case, with characteristics of both
confined and periodic system. We show that the resulting optical
spectrum correctly contains both the signature of defect levels located
within the band gap {\it and} electron-hole interactions. The latter
manifest themselves in the energy-resolved dielectric function as an
exciton feature below the electronic band gap. TDLDA is shown to
predict an optical transition between two defect levels, but the
exciton feature is not found. The paper is organized as follows: in
Section \ref{theory}, we present the theoretical framework, starting
from TDLDA and continuing through the GW method and solution of the
Bethe-Salpeter equation. Section \ref{applications} is devoted to
applications, containing one subsection for each particular system:
oligoacenes, silicon clusters, and F center in LiCl. We draw final
conclusions in Section \ref{conclusion}.

%%%%%%%%%%%%%%%%%%%%%%%%%%%%%%%%%%%%%%%%%%%%%%%%%%%%%%%%%%%%%%%%%%
%%%%%%%%%%%%%%%%%%%%%%%%%%%%%%%%%%%%%%%%%%%%%%%%%%%%%%%%%%%%%%%%%%
\newpage

\section{Theory}\label{theory}

\subsection{Linear response within TDLDA}\label{tdlda}

Gross and Kohn \cite{gross85} have shown how to extend DFT to the
time-dependent case by analyzing the effect on the charge density upon the
action of an external potential that changes in time. Their
theoretical approach can be further simplified by making use of two
assumptions \cite{onida02,vasiliev02,casida92}: adiabatic limit, and local-density
approximation, under which the exchange-correlation kernel is
instantaneous in time and a sole function of charge density at each
point in space. This is the time-dependent local-density approximation
(TDLDA) \cite{vasiliev02,casida92,casida98,jamorski95}.

When the external potential is due to an applied electric field,
the linear response in charge density $\rho$ is related to the
external potential via a response function (the polarizability)
according to

\begin{equation}
\Pi_f (1,2) = { \delta \rho (1) \over \delta V_{ext} (2) } \; \; ,
\label{e1.pi_rho}
\end{equation}
where we use a many-body notation for space-time and spin variables: $(1) =
({\bf r}_1,t_1, \tau_1)$. The subscript $f$ indicates that the
response function above is calculated within TDLDA. Working in frequency domain \cite{casida92}, the
TDLDA polarizability can be written as a sum over normal modes,

\begin{eqnarray}
\Pi_f (\rr , \rrp ; E) & = 2 \sum_s \rho_s(\rr ) \rho_s(\rrp ) \nonumber \\
& \times \left[ {1
    \over E - \omega_s + i0^+} - {1 \over E + \omega_s - i0^+} \right] ,
\label{e1.pi_pole}
\end{eqnarray}
where the normal modes of excitation in the system are denoted by
$\omega_s$, and $0^+$ represents a positive infinitesimal \footnote{As
  written, Eq. (\ref{e1.pi_pole}) corresponds to a time-ordered
  function, but it differs from the causal (measurable) response
  function only for negative values of energy.}. We assume
$\hbar =1$ throughout. The factor of 2 comes from summation over spin
indices. Still following the frequency-domain formulation by Casida
\cite{casida92,vasiliev02}, the amplitudes $\rho_s(\rr )$ are obtained
by solving a generalized eigenvalue problem. In order to obtain the
desired eigenvalue equation, we first expand $\rho_s$
in a series of single-particle transitions from an occupied level $v$
to an unoccupied level $c$:

\begin{equation}
\rho_s (\rr ) = \sum_{vc} X_{vc}^s \varphi_v(\rr ) \varphi_c(\rr )
\left( { \varepsilon_c - \varepsilon_v \over \omega_s} \right)^{1/2} \; \; ,
\label{e1.rho}
\end{equation}
where Kohn-Sham eigenvalues are denoted $\varepsilon_i$, with corresponding
eigenfunctions $\varphi_i$. For clarity, we reserve indices
$v$,$v'$ for occupied orbitals and indices $c$,$c'$
for unoccupied orbitals. The coefficients $X$ above satisfy an
eigenvalue equation in $(\omega^2)$ \cite{casida92}:

\begin{equation}
{\bf R}^{1/2} \left[ {\bf R} + 4 ( {\bf K}^x + {\bf K}^{LDA} ) \right]
{\bf R}^{1/2} {\bf X} = \omega_s^2 {\bf X} \; \; ,
\label{e1.tdlda}
\end{equation}
where ${\bf R}$, ${\bf K}^x$, and ${\bf K}^{LDA}$ are matrices in the
space of single-particle transitions:

\begin{equation}
R_{vcv'c'} = \delta_{vv'} \delta_{cc'} \left[ \varepsilon_c -
  \varepsilon_v \right] \; \; ,
\label{e1.rtdlda}
\end{equation}

\begin{equation}
K^x_{vcv'c'} = \int \dd \rr \int \dd \rrp
\varphi_v (\rr ) \varphi_c(\rr )
\vcoul (\rr , \rrp )
\varphi_{v'} (\rrp ) \varphi_{c'} (\rrp )  \; \; ,
\label{e1.kx}
\end{equation}

\begin{equation}
K^{LDA}_{vcv'c'} = \int \dd \rr
\varphi_v (\rr ) \varphi_c(\rr )
f_{xc} (\rr )
\varphi_{v'} (\rr ) \varphi_{c'} (\rr )  \; \; .
\label{e1.klda}
\end{equation}

The generalized eigenvectors ${\bf X}$ are normalized so that ${\bf
  X}^s {\bf X}^{s'} = \delta_{ss'}$. Throughout this article,
eigenfunctions $\varphi$ are assumed to be 
real functions. Indeed,
  they can be made real if the Kohn-Sham Hamiltonian is real and the
  electronic system is confined. This assumption is used in Eqs.
(\ref{e1.pi_pole}), (\ref{e1.rho}) and (\ref{e1.tdlda}). A notable situation when the equations above should be
  modified is in periodic systems, when eigenfunctions are often
  expressed as Bloch functions. The generalization to complex
  eigenfunctions is known \cite{onida02,reining02}. Also, we assume that the
  system has an energy gap and it is spin-unpolarized, although this
  is not an essential assumption and the same framework holds for
  gap-less systems ({\it e.g}, the F center in LiCl, Section
  \ref{licl}) as well. There
are extensions of TDLDA  to spin-polarized and/or gap-less systems \cite{casida92,vasiliev02,jamorski95}. Exchange-correlation
  effects are included in the kernel $f_{xc} = {\delta V_{xc} \over
  \delta \rho}$, which is a local, energy-independent quantity within
  TDLDA. The Coulomb potential is simply $\vcoul (\rr , \rrp ) = { e^2 \over |
  \rr - \rrp |}$.

Once the eigenvalue problem is solved, one can easily compute the
electrostatic polarizability tensor,

\begin{equation}
\alpha_{\beta \gamma} = {e^2 \over m} \sum_s { 
{\cal F}_s^{\beta \gamma} \over \omega^2_s } \; \; ,
\label{e1.alpha}
\end{equation}
where $m$ is the electron mass and ${\cal F}_s^{\beta \gamma}$ is the oscillator
strength along the three cartesian components $\beta,\gamma = \left\{ x, y, z \right\}$,

\begin{equation}
{\cal F}_s^{\beta \gamma} = 4 m \omega_s
\left( \int \dd \rr \rho_s (\rr ) \beta \right)
\left( \int \dd \rr \rho_s (\rr ) \gamma \right) \; \; .
\label{e1.ostdlda}
\end{equation}
By construction, the TDLDA polarizability
satisfies the oscillator-strength sum rule, 

\begin{equation}
\sum_s {\cal F}_s^{\beta \beta} = (\mbox{number of valence electrons} ) \; \; .
\label{e1.sumrule}
\end{equation}

Another quantity of interest is the
cross section for absorption of light in a confined system:

\begin{equation}
\sigma(E) = {2 \pi^2 e^2 \over m c} \sum_s {1 \over 3} \sum_\beta
{\cal F}_s^{\beta \beta} \delta(E - \omega_s) \; \; .
\label{e1.across}
\end{equation}

Fundamental quantities in linear-response theory are the
(longitudinal) dielectric function $\epsilon$ and its inverse
$\epsilon^{-1}$ \cite{strinati88,hedin69}. In order to discuss them,
we remind ourselves of the Kohn-Sham system: consider a set of
non-interacting electrons subject to a Hartree potential and an
exchange-correlation potential $V_{xc}$, chosen so that the
charge density of this fictitious system and of the real system are
identical \cite{martin,kohn65}. The presence of an external potential
induces charge redistribution and polarization in the Kohn-Sham system, which results in partial
screening of the applied potential $\delta V_{ext}$. For a
time-varying external perturbation, electrons are subject to an effective perturbation
potential given by

\[
\delta V_{eff} [\rho] (1) = \delta V_{ext} (1) + \delta V_{SCF} [\rho]
(1) \; \; ,
\]
where the self-consistent field is affected indirectly via charge
density, $\delta V_{SCF} = \vcoul \delta \rho + {\delta V_{xc} \over
  \delta \rho} \delta \rho $ \cite{casida92}. Defining the inverse
dielectric function as the change in effective potential due to an
external perturbation \cite{strinati88,hedin69},

\[
\epsilon^{-1} (1,2) = { \delta V_{eff} (1) \over \delta V_{ext} (2) } \; \; ,
\]
we then obtain a relationship between inverse dielectric function and
polarizability:

\[
\epsilon^{-1} (1,2) = \delta(1,2) + \int (3) \dd (3) \left[ \vcoul (1,3) +
  {\delta V_{xc} (1) \over \delta \rho (3)}  \right] \Pi (3,2) \; \; .
\]

In frequency representation, and using TDLDA for the
exchange-correlation potential $V_{xc}$, the above equation reduces to

\begin{eqnarray}
\epsilon_f^{-1}(\rr ,\rrp ; E) & = \delta(\rr ,\rrp ) + \int \dd \rrpp
\left[ \vcoul (\rr , \rrpp ) \right. \nonumber \\
& 
 + \left. f_{xc} (\rr ) \delta(\rr - \rrpp )
  \right] \Pi_f(\rrpp , \rrp ; E) \; \; .
\label{e1.epsinv}
\end{eqnarray}

Along similar lines, the dielectric function $\epsilon$ is given by

\begin{eqnarray}
\epsilon_f(\rr ,\rrp ; E) & = \delta(\rr ,\rrp ) - \int \dd \rrpp
\left[ \vcoul (\rr , \rrpp ) \right. \nonumber \\
&  \left. + f_{xc} (\rr ) \delta(\rr - \rrpp )
  \right] \chi_0(\rrpp , \rrp ; E) \; \; ,
\label{e1.eps}
\end{eqnarray}
where $\chi_0$ is the irreducible polarizability operator, within
the random-phase approximation (RPA) \cite{strinati88,adler62}:

\begin{eqnarray}
\chi_0 (\rr , \rrp ; E)  & = 2 \sum_s \varphi_v (\rr ) \varphi_c (\rr ) \varphi_v (\rrp ) \varphi_c (\rrp )  \nonumber \\
& \times \left[ {1
    \over E - \varepsilon_c + \varepsilon_v + i0^+} - {1 \over E + \varepsilon_c - \varepsilon_v - i0^+} \right] \; \; .
\label{e1.rpa}
\end{eqnarray}

Eqs. (\ref{e1.epsinv}) and (\ref{e1.eps}) describe screening of
the external scalar field {\it and} of the induced field. Recently,
Eq. (\ref{e1.eps}) was used to calculate the TDLDA static
screening in silicon clusters from first principles \cite{ogut03}.
The importance of Eq.
  (\ref{e1.epsinv}) is that it provides a direct relationship between
  polarizability and inverse dielectric function. For completeness, we
  present expressions for the dielectric function and its inverse
  within the RPA:

\begin{equation}
\epsilon_0^{-1}(\rr ,\rrp ; E) = \delta(\rr ,\rrp ) + \int \dd \rrpp
\vcoul (\rr , \rrpp ) \Pi_0(\rrpp , \rrp ; E) \; \; ,
\label{e1.epsinv0}
\end{equation}

\begin{equation}
\epsilon_0(\rr ,\rrp ; E) = \delta(\rr ,\rrp ) - \int \dd \rrpp
\vcoul (\rr , \rrpp ) \chi_0(\rrpp , \rrp ; E) \; \; ,
\label{e1.eps0}
\end{equation}
where $\Pi_0$ is the RPA polarizability, evaluated from
Eqs. (\ref{e1.pi_pole}), (\ref{e1.rho}) and (\ref{e1.tdlda})
after setting $f_{xc} = 0$.

%%%%%%%%%%%%%%%%%%%%%%%%%%%%%%%%%%%%%%%%%%%%%%%%%%%%%%%%%%%%%%%%%%
%%%%%%%%%%%%%%%%%%%%%%%%%%%%%%%%%%%%%%%%%%%%%%%%%%%%%%%%%%%%%%%%%%
\subsection{Electronic self-energy}\label{sigma}

A key quantity in solving the Bethe-Salpeter equation for optical
excitations is the electronic self-energy $\Sigma$, which can be computed from
first principles within the GW approximation (GWA)
\cite{hedin69,hybertsen86,aulbur00}. The basis of this approximation
lies in the so-called ``Hedin equations'', a set of non-linear
many-body equations which relate the self-energy with Green's function
$G$, polarizability $\chi$, screened Coulomb potential $W$, and vertex
function $\Gamma$ \cite{hedin69}:

\begin{equation}
W(1,2) = \vcoul (1,2) + \int \dd (34) \vcoul (1,3) \chi (3,4) W(4,2) \; \; ,
\label{e2.hedinw}
\end{equation}

\begin{equation}
\chi (1,2) = -i \int \dd (34) G(1,3) G(4,1^+) \Gamma (3,4;2) \; \; ,
\label{e2.hedinchi}
\end{equation}

\begin{equation}
\Sigma(1,2) = i \int \dd (34) G(1,3) W(4,1^+) \Gamma(3,2;4) \; \; ,
\label{e2.hedinsigma}
\end{equation}

\begin{eqnarray}
\Gamma(1,2;3) & = \delta(1,2) \delta(1,3) \nonumber \\
&  + \int \dd (4567) {\delta
  \Sigma (1,2) \over \delta G(4,5) } G(4,6) G(7,5) \Gamma (6,7;3)  \; \; .
\label{e2.hedingamma}
\end{eqnarray}

The approach taken by Hedin essentially generates a perturbation
series in the screened interaction $W$. This expansion is expected to
converge faster than an expansion in powers of the bare Coulomb
potential $V$, as long as electronic screening is strong. Following
this assumption \cite{hedin69}, the self-energy in
Eq. (\ref{e2.hedingamma}) is first taken as zero and the
vertex function reduces then to a delta function:

\[
\Gamma (1,2;3) \approx \delta(1,2) \delta(1,3) \; \; .
\]
With that vertex function, the polarizability $\chi$ reduces to the
RPA \cite{hedin69}:

\[
\chi(1,2) \approx -i G (1,2^+) G(2,1) \equiv \chi_0 (1,2) \; \; .
\]
The screened Coulomb interaction is evaluated in terms of the dielectric
function. From Eq. (\ref{e2.hedinw}), one gets:

\[
\epsilon_0 (1,2) = \delta(1,2) - \int \dd (3) \vcoul (1,3) \chi_0 (3,2) \; \; ,
\]

\begin{eqnarray*}
W_0 (1,2) & = \int \dd (3) \epsilon_0^{-1} (1,3) \vcoul (3,2) \\
&  = V (1,2) + \int \dd
(34) \vcoul (1,3) \Pi_0 (3,4) \vcoul (4,2) \; \; ,
\end{eqnarray*}
and the self-energy is

\begin{equation}
\Sigma (1,2) = i G (1,2) W_0 (2,1^+) \; \; .
\label{e2.gw0}
\end{equation}

The first {\it ab initio} calculations of self-energies for
semiconductors \cite{hybertsen86,godby88} followed the method outlined
above. For periodic systems, the dielectric function can be
conveniently expanded in plane wave basis and numerically
inverted. Matrix inversion is a reasonably inexpensive step when the
plane-wave expansion has a few hundreds or thousands of
components. Dynamical effects can be included for instance by inverting the
dielectric function at some values of frequency and either performing
numerical integration over the frequency axis \cite{aulbur00,godby88},
or using a generalized plasmon pole model \cite{hybertsen86}.

On the other hand, numerical inversion of the dielectric matrix in
real space is problematic because of the size of the matrix. As an
example, a converged calculation for the ground state of the silane
molecule (SiH$_4$) requires a real-space grid containing about $10^5$
points \cite{ogut03}. Inverting the dielectric function in real space
would require inverting a dense matrix of size $10^5 \times 10^5$. Although straightforward, this task
requires an extremely large amount of floating point operations and
CPU memory. This is certainly an extreme situation, but it illustrates
the type of problems involved with straight matrix inversion. The same
issue is present in periodic systems with large periodic cells: since
the number of plane-wave components in a Fourier expansion is
proportional to the volume of the cell, direct matrix inversion of
$\epsilon$ is also numerically expensive.

Significant numerical simplification is achieved by working in the
representation of single-electron transitions \cite{shirley93}, instead of using
plane-wave or real-space representations for the dielectric function.
there are two major advantages in doing that: the space
of transitions is often much smaller than either the real space or
reciprocal space, leading to matrices of reduced size; and
integrations over frequency can be performed analytically since the
polarizabilities $\chi_0$ and $\Pi_0$ have known pole structure. In the
space of transitions, a matrix element of the self-energy between
Kohn-Sham orbitals $j$ and $j'$ is given schematically by:

\begin{equation}
\bra j | \Sigma (E') | j' \ket = \bra j | i \int {\dd E \over 2 \pi}
e^{-iE0^+} G (E'-E) \left[ \vcoul + \vcoul \Pi_0 (E) \vcoul
  \right] | j' \ket \; \; .
\label{e2.sigma0me}
\end{equation}
The integral above can be replaced with a sum over
poles below the real energy axis. Following Hedin
\cite{hedin69,hedin95}, we write it as a summation of two terms: a
bare exchange contribution $\Sigma_x$ and a correlation contribution
$\Sigma_c$,

\begin{equation}
\bra j | \Sigma_x | j' \ket = - \sum_{n}^{occ.} K_{njnj'}^x \; \; ,
\label{e2.sigmax}
\end{equation}

\begin{equation}
\bra j | \Sigma_c (E) | j' \ket = 2 \sum_n \sum_s { V_{nj}^s V_{nj'}^s
  \over E - \varepsilon_n - \omega_s \eta_n } \; \; ,
\label{e2.sigmac}
\end{equation}
where
\begin{equation}
V_{nj}^s = \sum_{vc} K_{njvc}^x \left( {\varepsilon_c - \varepsilon_v
  \over \omega_s} \right)^{1/2} X_{vc}^s \; \; .
\label{e2.vpot}
\end{equation}

The derivation of Eqs. (\ref{e2.sigmax}), (\ref{e2.sigmac}) and
(\ref{e2.vpot}) from Eq. (\ref{e2.sigma0me}) is outlined in Appendix
\ref{appendix_integral}.
Whereas the summation over $n$ in Eq. (\ref{e2.sigmax}) is
performed over all occupied Kohn-Sham orbitals, Eq.
(\ref{e2.sigmac}) has a summation over occupied and unoccupied
orbitals. One can accelerate convergence in the last summation by
truncating it and
evaluating the remainder under the static approximation (see Appendix
\ref{appendix_static}). The coefficient
$\eta_n$ has value +1 for unoccupied orbitals and -1 for occupied
orbitals. In the equations above, eigenvectors $X$ and
eigenvalues $\omega$ are still evaluated under the random-phase
approximation, obtained by setting $f_{xc}=0$ in Eq. (\ref{e1.tdlda}).
In the following discussion, we refer to this level of approximation for the
self-energy as GW$_0$.

At this point, it is natural to advance one step further and use
dielectric screening within TDLDA instead of RPA. The impact of TDLDA
screening in the context of the GW approximation in bulk silicon has
been analyzed before \cite{hybertsen86,delsole94,bruneval05}. In order to have a
controlled level of approximation in Hedin's equations, we return to
Eq. (\ref{e2.hedingamma}) and start an iterative solution by assuming
$\Sigma(1,2) \approx V_{xc} (1) \delta(1,2)$. The vertex function then
becomes

\begin{eqnarray}
\Gamma(1,2;3) & \approx \delta(1,2) \delta(2,3) + \int \dd (45) \times \nonumber \\
&  \left[
  -i \delta(1,2) f_{xc} (1) \right] G(1,4) G(5,1^+) \Gamma(4,5;3) \; \; .
\label{e2.gamma}
\end{eqnarray}
The irreducible polarizability $\chi$ is now

\begin{equation}
\chi(1,2) = \chi_0(1,2) + \int \dd (3) \chi_0 (1,3) f_{xc} \chi (3,2) \; \; .
\label{e2.chi}
\end{equation}

For the screened Coulomb interaction, we write Eq.
(\ref{e2.hedinw}) in terms of the full polarizability operator:

\[
W(1,2) = \vcoul (1,2) + \int \dd (34) \vcoul (1,3) \Pi (3,4)
\vcoul (4,2) \; \; ,
\]
with
\[
\Pi(1,2) = \chi(1,2) + \int \dd (34) \chi(1,3) \vcoul (3,4) \Pi(4,2) \; \; .
\]
From Eq. (\ref{e2.chi}), we get

\begin{eqnarray*}
\Pi(1,2) & = \chi_0 (1,2) + \int \dd (34) \chi_0 (1,3) \left[ \vcoul
  (3,4) \right. \nonumber \\
& + \left. f_{xc} (3) \delta(3,4) \right] \Pi(4,2) \; \; .
\end{eqnarray*}

Although written in real-space representation, the function $\Pi$
above is identical to Eq. (\ref{e1.pi_pole}), and we refer to it
as $\Pi_f$ in the subsequent discussion. This function describes
polarization due to an external potential accompanied by dynamical
screening  produced by the self-consistent field.
Finally, we arrive at the following expression
for the self-energy:

\begin{eqnarray}
\Sigma(1,2) & = i G(1,2) \left[ \vcoul (1^+,2) + \int \dd (34) \left\{
  \vcoul (1,3) \right. \right. \nonumber \\
& + \left. \left. f_{xc} (1) \delta(1,3) \right\} \Pi_f (3,4) \vcoul
  (4,2) \right] \; \; .
\label{e2.gwf}
\end{eqnarray}

We note that the use of TDLDA screening
in the self-energy causes the inclusion of a vertex term (the second
term inside curly brackets above). This level of approximation has been
used before in the study of the quasi-particle band structure of
crystalline silicon \cite{delsole94}.
Eq. (\ref{e2.gwf}) is a valid approximation for the self-energy but it is
not symmetric with respect to the interchange of indices 1 and 2. In
principle, this symmetry is not present in Eq.
(\ref{e2.hedinsigma}), but it can be recovered by defining a
``left-sided'' vertex function written schematically as $\Sigma = i
\Gamma W G$ as opposed to
$\Sigma = i G W \Gamma $ ({\it c.f.} Eq. \ref{e2.hedinsigma})
\cite{strinati88}. This deficiency is corrected by symmetrizing the
last term in Eq. (\ref{e2.gwf}). Schematically, we rewrite
$\Sigma = i G [ V + V \Pi_f V + f \Pi_f V ]$ as $\Sigma = i G [ V + V \Pi_f
  V + {1 \over 2} V \Pi_f f + {1 \over 2} f \Pi_f V ]$. Matrix
elements of the self-energy are then given by:

\begin{widetext}
\begin{figure}[ht]
\centering\epsfig{figure=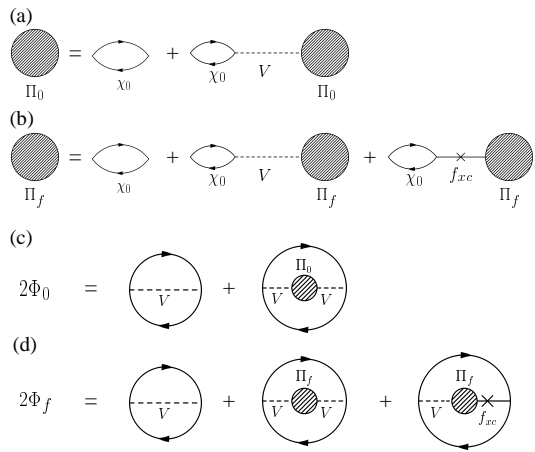,width=7cm,clip=}
\caption{Feynman diagrams for the polarizability
  operator $\Pi$ (a,b) and the function $\Phi$ (c,d). The local
  exchange-correlation kernel is represented by a crossed line. Solid
  oriented lines are Green's functions. Dashed lines denote the bare Coulomb
  potential.}
\label{f2.feynman}
\end{figure}

\begin{equation}
\bra j | \Sigma (E') | j' \ket = \bra j | i \int {\dd E \over 2 \pi}
e^{-iE0^+} G (E'-E) \left[ \vcoul + \vcoul \Pi_f (E) \vcoul + {1
 \over 2} \vcoul \Pi_f (E) f_{xc} + {1 \over 2} f_{xc} \Pi_f (E) \vcoul
  \right] | j' \ket \; \; .
\label{e2.sigmame}
\end{equation}
\end{widetext}

This self-energy operator is written as a sum of three
contributions: $\Sigma = \Sigma_x + \Sigma_c + \Sigma_f$, where the
first two terms are given by Eqs. (\ref{e2.sigmax}) and
(\ref{e2.sigmac}) respectively, but now with full TDLDA screening ($f_{xc}
\neq 0$). The last term is a vertex correction:

\begin{equation}
\bra j | \Sigma_f (E) | j' \ket = \sum_n \sum_s { V_{nj}^s F_{nj'}^s
+  F_{nj}^s V_{nj'}^s \over E - \varepsilon_n - \omega_s \eta_n } \; \; ,
\label{e2.sigmaf}
\end{equation}
with
\begin{equation}
F_{nj}^s = \sum_{vc} K_{njvc}^{LDA} \left( {\varepsilon_c - \varepsilon_v
  \over \omega_s} \right)^{1/2} X_{vc}^s \; \; .
\label{e2.fpot}
\end{equation}
In order to make distinction between the two levels of approximation,
we refer to the last approximation (Eq. \ref{e2.sigmame}) as GW$_f$,
as opposed to GW$_0$, Eq. (\ref{e2.sigma0me}).

It is not unexpected that, by using a polarizability function from
TDLDA, we obtain a self-energy operator that has vertex
corrections. In the language of many-body physics, the assumption in
Eq. (\ref{e2.gamma}) implies that additional Feynman diagrams are
included in the polarizability and in the self-energy. Fig.
\ref{f2.feynman}(a,b) shows the diagrams included in the polarizability
$\Pi$ for both RPA and TDLDA. Although such diagrams include the LDA
kernel, which does not have expansion in $G$ and $\vcoul $, we can
still define a many-body function $\Phi$ such that $\Sigma(1,2) =
\delta \Phi / \delta G(1,2)$. The diagrams for $\Phi$ are depicted in
Fig. \ref{f2.feynman}(c,d). Adding the last diagram in Fig.
\ref{f2.feynman}(b) amounts to enhanced screening, since it allows for a
new channel for electrons to redistribute in the presence of an
external potential. This is counterbalanced by  the inclusion of a
vertex diagram (right-most diagram in Fig. \ref{f2.feynman}(d)). We
also note that Fig. \ref{f2.feynman} provides a justification for
symmetrizing Eq. (\ref{e2.gwf}): differentiating the additional vertex
diagram with respect to G leads to a pair of diagrams consistent with
Eq. (\ref{e2.sigmaf}).

Inclusion of self-energy corrections in the description of the
electronic system is done following the usual procedure: we assume the
quasiparticle approximation and write down an eigenvalue equation for
quasiparticles \cite{hybertsen86,aulbur00},

\begin{equation}
\left[ H_{LDA} + \Sigma - V_{xc} \right] \psi_j = E_j \psi_j \; \; ,
\label{e2.hqp}
\end{equation}
where quasiparticle orbitals $\psi_j$ are expanded in the basis of
Kohn-Sham eigenfunctions and $H_{LDA}$ is the (diagonal) LDA
Hamiltonian \cite{aulbur00}. Quasiparticle energies and wave-functions
are now found by direct diagonalization of the equation above.

In writing Eq. (\ref{e2.hqp}), we should consider carefully the
energy dependence of $\Sigma$. Hybertsen and Louie \cite{hybertsen86}
have shown that evaluating the self-energy around the quasiparticle
energy leads to accurate band structures for cubic
semiconductors. Since the quasiparticle energy is not known before
Eq. (\ref{e2.hqp}) is solved, the suggested procedure \cite{hybertsen86} is to
evaluate the operator $\bra \Sigma (E) - V_{xc} \ket$ and its energy
derivative at the LDA eigenvalue, $E=E_{LDA}$, and use linear
extrapolation for the actual quasiparticle energy. We follow a similar
methodology: first, we evaluate the diagonal part of the quasiparticle
Hamiltonian in Eq. (\ref{e2.hqp}) and obtain a first estimate for
quasiparticle energies by solving the equation $E_j = \varepsilon_j +
\bra j | \Sigma (E_j) - V_{xc} | j \ket$. In the next step, we include
off-diagonal matrix elements and proceed through full
diagonalization. At the end, quasiparticle energies are still close to
their first estimate.

An open issue regarding the GW method is whether self-consistency
between self-energy, polarizability and Green's function should be
imposed or not and, if so, how to do it \cite{fleszar05}. When the GW$_0$ and GW$_f$
approximations were obtained by iterating Hedin's equations,
self-consistency was lost and the vertex function was drastically
simplified. Early work has indicated that self-consistency and
vertex contributions partially cancel each other \cite{dubois59}, which
may explain the remarkable success of the GW method for semiconductors. There
have been some attempts at imposing partial self-consistency between
self-energy and Green's function \cite{hybertsen86,aulbur00}, which
resulted in quasiparticle energies improved by a fraction of
electron-volt. Nevertheless, full self-consistency between self-energy
and Green's function was observed to degrade the quasiparticle
bandwidth and the description of the satellite structure in the
electron gas \cite{holm98}.
In the subsequent applications, we do not attempt to
impose self-consistency. Instead, the single-particle Green's function
is always constructed from Kohn-Sham eigenvalues and eigenfunctions.

%%%%%%%%%%%%%%%%%%%%%%%%%%%%%%%%%%%%%%%%%%%%%%%%%%%%%%%%%%%%%%%%%%
%%%%%%%%%%%%%%%%%%%%%%%%%%%%%%%%%%%%%%%%%%%%%%%%%%%%%%%%%%%%%%%%%%

\subsection{Bethe-Salpeter equation}\label{bse}

While the GW method provides a description of quasiparticles in the
system ({\it i.e.}, the energy needed to add or extract one electron
from the system), the solution of the Bethe-Salpeter equation gives information
about neutral excitations: the process of promoting electrons from
occupied quasiparticle orbitals to unoccupied ones. The important
Green's function 
now is no longer the one-electron Green's function but the
two-particle or, more specifically, the electron-hole Green's function
\cite{strinati88}.

We start with a many-body expression for the polarizability:

\begin{equation}
\Pi (1,2) = -i L (1,2; 1^+,2^+) \; \; ,
\label{e3.pi}
\end{equation}
where $L$ is the electron-hole correlation function
\cite{onida02,strinati88,rohlfing00}, which satisfies the Bethe-Salpeter
equation:

\begin{eqnarray}
& L(1,2;3,4) = G(1,4) G(2,3) \nonumber \\
& + \int \dd (5678) G(1,5) G(6,3) K(5,7;6,8)
L(8,2;7,4) \; \; .
\label{e3.bsel}
\end{eqnarray}

The kernel operator $K$ describes interactions between the excited
electron and the ``hole'' left behind in the electron sea. The
connection with GW is present in two aspects:

\begin{enumerate}
\item The Green's function that enters in Eq. (\ref{e3.bsel}) is
  the Green's function for the interacting electronic system. Within
  the quasiparticle approximation, it is calculated with eigenvalues and
  wave-functions obtained by diagonalizing Eq. (\ref{e2.hqp}).
\item The kernel $K$ is related to the self-energy by
  \cite{onida02,strinati88,rohlfing00}:

\begin{equation}
K(1,2;3,4) = -i \delta(1,3) \delta(2,4) \vcoul (1,2) + {\delta \Sigma
  (1,3) \over \delta G(4,2) } \; \; .
\label{e3.kernel1}
\end{equation}

\end{enumerate}

Although the functions $G$ and $K$ can be evaluated in many
approximations, it is important to retain the same level of
approximation in both quantities, so that all important Feynman
diagrams are included and no diagrams are double-counted. In the lowest level of approximation, many-body
effects can be ignored completely and the self-energy approximated by
the LDA exchange-correlation potential, $\Sigma(1,2) \approx V_{xc}
(1) \delta(1,2)$. In this case, the Green's function to be used is
constructed from LDA eigenvalues and eigenfunctions, and Eq.
(\ref{e3.bsel}) reduces itself to the TDLDA eigenvalue equation,
Eq. (\ref{e1.tdlda}). This fact has been explored in recent
studies where model exchange-correlation functionals are designed to
work as good approximations for the self-energy in Eq.
(\ref{e3.kernel1}) while retaining as much as possible the formal
simplicity of TDLDA \cite{onida02,reining02,marini03}. By its own nature, this
family of approximations does not have Feynman diagrams in terms of
which the self-energy is expanded.

Another level of approximation is GW$_0$, Eq. (\ref{e2.sigma0me}). The
kernel $K$ has then two terms: a bare exchange interaction, originated from
the first term in Eq. (\ref{e3.kernel1}), and a screened
interaction, from $\delta \Sigma / \delta G = i W_0$. This
approximation has been used in a variety of different applications
with very good results (see {\it e.g.} \cite{onida02,rohlfing00}).

Finally, a third level of approximation is GW$_f$, Eq.
(\ref{e2.sigmame}). At this level, the kernel has an
additional term besides the previous two: a LDA vertex correction. As
opposed to the kernel present in the TDLDA equation, the BSE kernel
within either GW$_0$ or GW$_f$ has explicit energy dependence, from
the polarizability operator $\Pi$. This dependence was observed to be
negligible when the interaction kernel itself is weak
\cite{rohlfing00}. Ignoring dynamical effects and ignoring the mixing
of absorption and emission contributions in the kernel 
\footnote{
Confirming earlier work \cite{rohlfing00}, we have observed that the
Tamm-Dancoff approximation ({\it i.e.}, neglecting mixing between
absorption and emission contributions) is acceptable when calculating
excitation energies in the GW+BSE framework. This contrasts with the
scenario in TDDFT, where mixing is usually important
\cite{onida02,casida92,furche01}, due to the specific properties of
the TDDFT kernels. The impact of the Tamm-Dancoff approximation in the
energy loss spectrum of silicon has been analyzed by Olevano and Reining
\cite{olevano01}.
}, one can rewrite Eq. (\ref{e3.bsel}) as an eigenvalue equation
\cite{onida02,rohlfing00}:

\begin{eqnarray}
& \Omega_l A^l_{vc}  = \left( E_c - E_v \right) A^l_{vc} \nonumber \\
&  + \sum_{v'c'} \left( 2 K^x_{vcv'c'}
+ K^d_{vcv'c'} + K^f_{vcv'c'} \right) A^l_{v'c'} \; \; ,
\label{e3.bse}
\end{eqnarray}
where $E_c,E_v$ are quasiparticle energies of unoccupied and
occupied orbitals respectively. The exchange kernel $K^x$ has the same
form given by Eq. (\ref{e1.kx}). The kernels $K^d$ and $K^f$ are
given by:

\begin{equation}
K^d_{vcv'c'} = \ K^x_{vv'cc'} + 4 \sum_s { V^s_{vv'} V^s_{cc'} \over
  \omega_s} \; \; ,
\label{e3.kerneld}
\end{equation}

\begin{equation}
K^f_{vcv'c'} = 2 \sum_s { V^s_{vv'} F^s_{cc'} + F^s_{vv'} V^s_{cc'} \over
  \omega_s} \; \; .
\label{e3.kernelf}
\end{equation}

The eigenvectors $A$ are normalized in the usual way, and the
polarizability is now given by

\begin{equation}
\Pi_{BSE} (\rr , \rrp ; E) = \sum_l \left[ { \rho_l(\rr ) \rho_l (\rrp )
    \over E - \Omega_l + i0^+} - { \rho_l(\rr ) \rho_l (\rrp )
    \over E + \Omega_l - i0^+} \right] \; \; ,
\label{e3.pibsepole}
\end{equation}
with

\begin{equation}
\rho_l (\rr ) = \sum_{vc} \varphi_v (\rr ) \varphi_c (\rr ) A^s_{vc} \; \; .
\label{e3.bserho}
\end{equation}

As in the TDLDA, the solution of the Bethe-Salpeter equation provides
the electrostatic susceptibility tensor and absorption cross section
by equations analogous to
(\ref{e1.alpha}),(\ref{e1.ostdlda}),(\ref{e1.across}), with the
replacements $\rho_s \rightarrow \rho_l$ and $\omega_s \rightarrow
\Omega_l$.

%%%%%%%%%%%%%%%%%%%%%%%%%%%%%%%%%%%%%%%%%%%%%%%%%%%%%%%%%%%%%%%%%%
%%%%%%%%%%%%%%%%%%%%%%%%%%%%%%%%%%%%%%%%%%%%%%%%%%%%%%%%%%%%%%%%%%

\subsection{Technical considerations}\label{numerical}

Working in the space of single-particle transitions has two clear
advantages: it does not assume any particular boundary condition over
electron wave-functions, being equally valid for confined as well as
extended systems; and it often leads to matrices smaller than the
corresponding ones in real-space or reciprocal-space representations.
In addition, the necessary
numerical tasks are simple: matrix algebra (diagonalizations and
matrix products), and evaluation of integrals of the type in Eqs.
(\ref{e1.kx}) and (\ref{e1.klda}). We can also take advantage of
existing point symmetries in the system at hand and divide the
transition space into different group representations, resulting in
block-diagonal matrices. This method can be readily extended to periodic
systems by adding lattice periodicity as an additional symmetry
operation. Each block is then associated to a particular k-point in the
Brillouin zone. For large systems, the solution of Eqs.
(\ref{e1.tdlda}) and (\ref{e3.bse}) may involve large matrices, but
further simplification is obtained by restricting the transition space
to low-energy transitions. In the calculation of the self-energy terms
$\Sigma_c$ and $\Sigma_f$, one can proceed through the sum over
single-particle states by including the static remainder (see appendix
\ref{appendix_static}).

%%%%%%%%%%%%%%%%%%%%%%%%%%%%%%%%%%%%%%%%%%%%%%%%%%%%%%%%%%%%%%%%%%
%%%%%%%%%%%%%%%%%%%%%%%%%%%%%%%%%%%%%%%%%%%%%%%%%%%%%%%%%%%%%%%%%%

\section{Applications}\label{applications}

In order to assess the reliability of the current implementation, we
have employed it to study the excited-state properties of different
classes of systems, both localized and extended. In all applications,
we start from the electronic system in its ground state, as described within
DFT. The exchange-correlation potential is used in the Ceperley-Alder
form \cite{perdew81}. With the use of norm-conserving
pseudopotentials \cite{troullier90}, we are able to remove core
electrons from the problem, reducing considerably the numerical
effort. In all applications, we use pseudopotentials constructed
according to the Troullier-Martins prescription
\cite{troullier90}. We employ a real-space
approach in the solution of the Kohn-Sham
Equation \cite{chelikowsky94}. In this approach, wave-functions are expressed
directly as functions of position. For a finite system, wave-functions
are required to vanish outside a spherical boundary with adjustable
radius. The volume inside this boundary is
sampled by a homogeneous grid with fixed grid spacing $h$ along the
three Cartesian directions. Results are tested for convergence with
respect to the grid spacing and the boundary radius. We explore point
symmetries both in the solution of the Kohn-Sham equation and in the
calculation of integrals in Eqs. (\ref{e1.kx}) and (\ref{e1.klda}). For the self-energy operator, we include the
static remainder according to appendix \ref{appendix_static}. We
carefully test convergence with respect to the highest state
explicitly included in the $n$ summation. Except when noted, we
compute self-energies at the GW$_f$ level. Numerical accuracy in
self-energy matrix elements is 0.2~eV or better. Excited states of the
electronic system are calculated without the inclusion of effects due
to structural relaxation.

\subsection{Benzene and naphthalene}\label{oligoacenes}

Having a relatively small number of valence electrons makes the
benzene molecule, C$_6$H$_6$, a good test system for the theory just
presented. In addition, benzene has been the subject of extensive
experiments, probing various properties such as linear optical response,
electronic structure and ionization \cite{nist}. On the theory side,
its optical properties have been investigated with both
TDLDA and GW-BSE approaches
\cite{vasiliev02,tiago05,bertsch01}. Quasiparticle energy levels of
benzene and other oligoacenes have also been recently calculated using
a first-principles Gaussian orbital-based method as well as
a tight-binding approach \cite{niehaus05}. The series of oligoacenes,
benzene being the first element, is a sequence of molecules composed
of $n$ co-joined aromatic rings, starting from $n=1$ (benzene), $n=2$
(naphthalene), {\it etc}. Oligoacene crystals have received renewed attention
recently due to their potential use in organic thin film devices
\cite{tiago03,park02}.

\begin{figure}[t]
\centering\epsfig{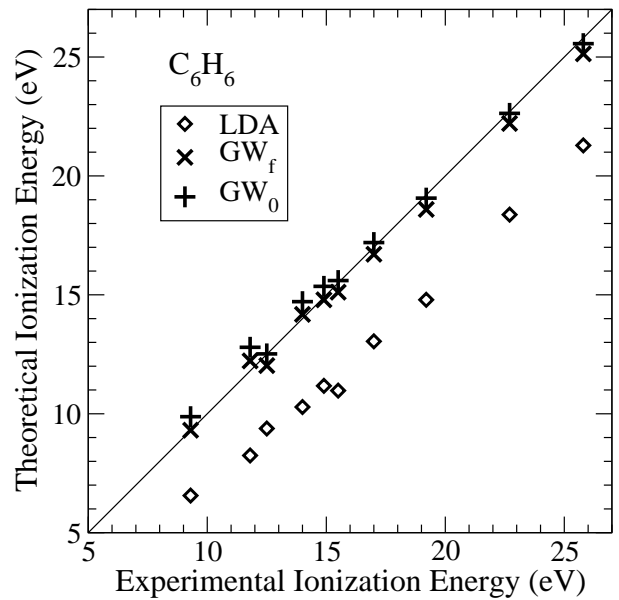}
\caption{Ionization energies of benzene,
  associated to all occupied orbitals in the molecule, as predicted
  within GW$_0$ and GW$_f$. The theoretical ionization energy is the negative of
  a quasiparticle energy eigenvalue. Experimental data from \cite{lipari76}.}
\label{f4.benzeneqp}
\end{figure}

In our calculations, we solve the Kohn-Sham equations on a regular
grid with grid spacing 0.4~a.u. (the Bohr radius being 1~a.u.). Electronic wave-functions were required to vanish outside a
sphere of radius 16~a.u. centered on the molecule. Carbon-carbon and
carbon-hydrogen bond lengths were fixed at their experimental
values. The calculation of the self-energy operator was done including
up to 305 unoccupied (virtual) orbitals in the TDLDA polarizability
and in the calculation of correlation and vertex parts. This
ensures that the sum rule is satisfied to within 30\%, which was found
to be sufficient for an accuracy of 0.1~eV in quasiparticle
energies. After the self-energy operator was computed, quasiparticle
energies for occupied and unoccupied orbitals were obtained by
diagonalizing the Hamiltonian in Eq. (\ref{e2.hqp}).
Extensive convergence tests have been performed on all relevant
numerical parameters: choices of 0.3 and 0.4~a.u. were used for the
grid spacing; we have used boundary radii of up to 20~a.u., and
included up to 1000 virtual orbitals. For the last choice, the sum
rule is satisfied to within 14\%.

Fig. \ref{f4.benzeneqp} shows a comparison between theoretical and
experimental ionization energies. In the framework of the GW
approximation, the first ionization energy is interpreted as the
negative of the quasiparticle energy for the highest occupied
molecular orbital (HOMO). Higher ionization energies are obtained from
deeper molecular orbitals. As reference, we show in Fig.
\ref{f4.benzeneqp} the prediction from LDA ({\it i.e.}, negative of
the Kohn-Sham eigenvalues for occupied orbitals), although such comparison
should be made with caution since, from a strict point of view,
such eigenvalues enter in DFT as mathematical entities, with no
explicit physical meaning. 
By extending Koopman's theorem to DFT, the HOMO eigenvalue can be
shown to be equal to the first ionization potential if the exact
functional is used \cite{krieger92}, but such property does not hold
for higher ionization potentials \cite{krieger92,janak78}. Hybrid
functionals \cite{zhan03} and the ones of the exact-exchange type
\cite{dellasala01,gorling99_1} can give very good predictions for the
first ionization potential.

In contrast, quasiparticle energies do have physical meaning, and
Fig. \ref{f4.benzeneqp} shows definitive consistency between theory
and experiment. In particular, the first ionization energy is
predicted to be 9.30~eV within GW$_f$, with remarkable agreement with
the experimental value of 9.3~eV \cite{lipari76,nist} 
and also with the prediction from quantum chemistry
calculations: 9.04~eV \cite{deleuze03}. A hybrid B3LYP functional predict
9.74~eV for this ionization energy \cite{hirata03}.
The GW$_0$ approximation predicts the same quantity to be 9.88~eV. The
difference is due mostly to the vertex term: the contribution
$\Sigma_f$ for the HOMO alone is found to be almost 0.8~eV, while LDA
exchange-correlation corrections in the polarizability give a
contribution somewhat smaller but of opposite sign. For other
quasiparticle energies, the vertex term is always positive
({\it i.e.}, it lowers the ionization energy compared to the GW$_0$
value), and no more than 1~eV.

Ionization energies in the range of 20 to 25~eV are found to be
systematically underestimated with respect to experiment, as shown in
Fig. \ref{f4.benzeneqp}. For such deep orbitals, we expect to have
some loss of numerical accuracy since, although the energy dependence
of the polarizability $\Pi_f$ is exactly calculated using Eq.
(\ref{e1.pi_pole}), the summation over $s$ is always limited to a
finite number of poles.

\begin{table}[b]
\caption{Electron affinities in benzene, as predicted
  within GW$_0$, Eq. (\ref{e2.sigma0me}) and GW$_f$, Eq.
  (\ref{e2.sigmame}). 
The outer-valence Green's function (OVGF) method \cite{deleuze03} is
  based on a Hartree-Fock expansion of the self-energy, combined with
  self-consistent calculations of self-energy and Green's function.
All energies in eV.}
\label{t4.benzeneea}
\begin{center}
\begin{tabular}{ccccccc} \hline \hline
 & \hspace{0.cm} LDA
 & \hspace{0.cm} GW$_0$
 & \hspace{0.cm} GW$_f$
 & \hspace{0.cm} DFT(B3LYP) \cite{rienstra-kiracofe01}
 & \hspace{0.cm} OVGF \cite{deleuze03}
 & \hspace{0.cm} Exp.\cite{burrow87} \\ \hline
$e_{2u}$ & 1.33 & -0.47 & -0.99 & -0.88 & -2.605 & -1.12 \\
$b_{2g}$ & -2.45 & -4.49 & -5.05 & & & -4.82 \\
\hline \hline
\end{tabular}
\end{center}
\end{table}

\begin{figure}[t]
\centering\epsfig{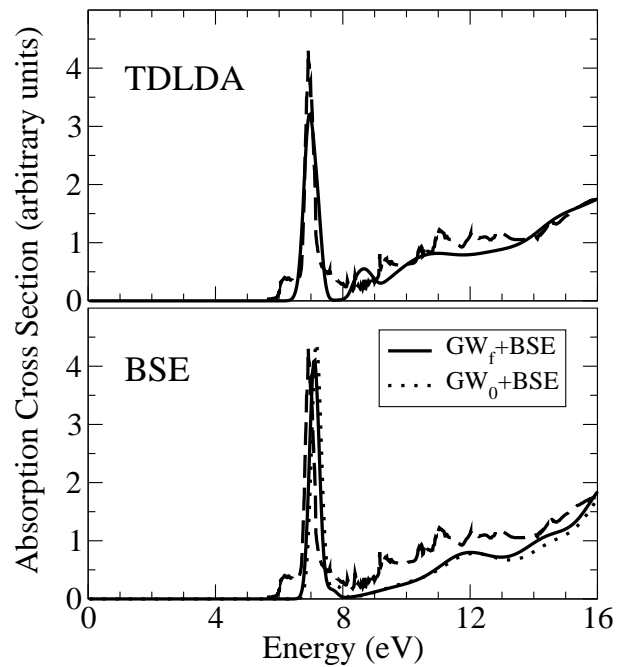}
\caption{Absorption spectrum of benzene,
  calculated within TDLDA (upper panel) and BSE (lower
  panel). Absorption lines were broadened by Gaussian convolution with
  dispersion 0.15~eV (below 10~eV) and 0.5~eV (above 10~eV). The
  measured spectrum  \cite{koch72} is shown in dashed
  lines. In the lower panel, the BSE was solved using two
  approximations for the self-energy: GW$_0$ (dotted line) and GW$_f$
  (solid line).}
\label{f4.benzenebse}
\end{figure}

The quasiparticle energy for the lowest unoccupied molecular orbital (LUMO) is
the negative of the electron affinity \cite{hedin69,rohlfing00}, and
its calculated value is shown in Table \ref{t4.benzeneea}. Although
the anion C$_6$H$_6^-$ is unstable, Burrow and collaborators
\cite{burrow87} have been able to perform careful electron
transmission experiments and identify resonances in the spectrum,
which where associated to the electron affinity. Two resonances where
found: at energies 1.12~eV and 4.82~eV. They match theoretical
predictions from the GW$_f$ approximation both in energy
and spacial distribution: $e_{2u} (\pi^*)$ and $b_{2g}(\pi^*)$
respectively. Again, there is a discrepancy between GW$_0$ and GW$_f$
quasiparticle energies of about 0.5~eV, mostly due to vertex
corrections in the self-energy.

\begin{table*}
\caption{Excitation energies of the lowest-energy neutral excitations
  in benzene. Energies in eV.}
\label{t4.benzenebse}
\begin{center}
\begin{tabular}{cccccccc} \hline \hline
  \hspace{0.cm}
 & \hspace{0.cm} TDLDA
 & \hspace{0.cm} GW$_0$+BSE
 & \hspace{0.cm} GW$_f$+BSE
 & \hspace{0.cm} TDDFT(B3LYP) \cite{heinze00}
 & \hspace{0.cm} TDDFT(LHFX) \cite{dellasala01}
 & \hspace{0.cm} CASPT2 \cite{lorentzon95}
 & \hspace{0.cm} Exp. \cite{doering69} \\ \hline
\multicolumn{2}{l}{Triplet} \\ 
 $B_{1u}$ & 4.53 & 3.59 & 3.59 & 4.45 & 4.27 & 3.89 & 3.9 \\ \hline
\multicolumn{2}{l}{Singlet} \\
 $B_{2u}$ & 5.40 & 4.74 & 4.86 & 5.24 & 5.40 & 4.84 & 5.0 \\
  $B_{1u}$ & 6.23 & 6.08 & 6.14 & 6.09 & 6.12 & 6.30 & 6.2 \\
  $E_{1u}$ & 6.9-7.2 & 7.16 & 7.23 & 6.90 & 6.96 & 7.03 & 6.9 \\
\hline \hline
\end{tabular}
\end{center}
 \end{table*}

If measuring electron affinities of benzene is not trivial, a similar
statement can also be made regarding first-principles calculations. The
anion C$_6$H$_6^-$ is correctly predicted within DFT to be unstable,
and methods such as $\Delta$SCF (energy variations in self-consistent
field) fail to predict its
electron affinity \cite{tiago05}. Indeed, confining the anion inside a spherical
boundary of radius $R$ and solving the Kohn-Sham equations results in
an excess energy relative to the neutral system that is proportional to
$R^{-2}$. This excess energy vanishes in the limit of very large
radius. The physical picture is that, although the anion is unstable,
it can be detected as a resonance if electrons do not remain in their
ground state, and the GW approximation predicts that resonance as a
quasiparticle orbital.

As mentioned in the context of Eq. (\ref{e2.hqp}), quasiparticle
wave-functions are obtained by numerically diagonalizing the
quasiparticle Hamiltonian, and therefore they are different from
Kohn-Sham eigenfunctions. For the benzene molecule, we observed though
that quasiparticle and Kohn-Sham eigenfunctions overlap by more than
95 \% for all occupied orbitals. This is due to the high symmetry of
the benzene molecule, which makes most off-diagonal matrix elements of
the self-energy zero by selection rules. Similarly, the resonant states $e_{2u}$
and $b_{2g}$ in Table \ref{t4.benzeneea} have overlaps of 99 \% and 63
\% respectively. For them, overlap is still suppresed by
selection rules but the existence of many low-energy, unbound orbitals
makes overlap somewhat more favorable for orbitals well above the
HOMO-LUMO gap.

Optical excitations were obtained in two methods: both within TDLDA
and by solving the BSE. Fig. \ref{f4.benzenebse} shows the cross
section for photoabsorption in benzene. It is dominated by a sharp and
well pronounced peak at around 7~eV, as verified by various calculations
  \cite{vasiliev02,bertsch01,vasiliev04}
\footnote{
The value of 7.4~eV reported in Ref. \cite{vasiliev02} for the peak
position has been recently revised to 7.1~eV in Ref. \cite{vasiliev04}.
}. This peak is the visible
component of a $\pi-\pi^*$ complex, originated from transitions between
the HOMO and the quasiparticle state $e_{2u}$. The low,
flat feature in the 6.0-7.0~eV range is due to coupling between
$\pi-\pi^\star$ transitions and
vibrational modes \cite{herzberg,bertsch01}, and it is absent in the
calculated spectra because of the assumed structural rigidity. Beyond
10~eV, a number of sharp features in the measured spectrum results from
transitions involving Rydberg states \cite{herzberg,koch72}. The
limited numerical accuracy in that energy range prevents a detailed
identification of such transitions in the calculated spectrum.

Table \ref{t4.benzenebse} shows a comparison between TDLDA and BSE predictions
for some excitations in the $\pi-\pi^\star$ complex, 
together with CASPT2 calculations \cite{lorentzon95} and other TDDFT
calculations \cite{dellasala01,heinze00}. Although singlet
transitions $B_{1u}^1$ and $E_{1u}^1$, which are the dominant ones in the
low-energy part of the absorption spectrum, are equally well described 
by both methods, there
is a significant blue shift of dark transitions $B_{1u}^3$
and $B_{2u}^1$ within all TDDFT approaches presented.
With the exception of the $E_{1u}^1$
transition, excitation energies obtained by solving
the BSE are typically underestimated with respected to measured
quantities. The overall deviation between measured 
transition energies and GW predictions is 0.1 to 0.3~eV, 
comparable to CASPT2 results \cite{lorentzon95}.

\begin{figure}[b]
\centering\epsfig{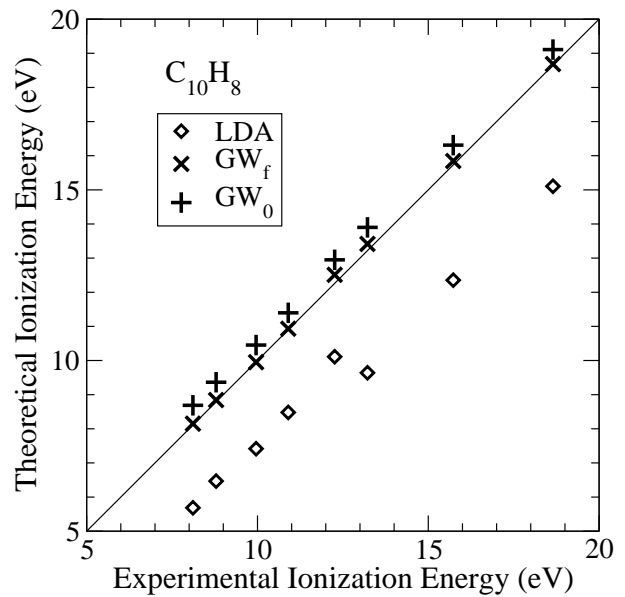}
\caption{Ionization energies of naphthalene,
  associated to all occupied $\pi$ orbitals and the first $\sigma$
  orbitals in the molecule. Experimental data from \cite{dewar69}.}
\label{f4.nleneqp}
\end{figure}

Although excited states can be obtained from either TDLDA or BSE by
solving suitable eigenvalue problems, these two theories provide very
different descriptions for ``electron-hole'' correlations. Within
TDLDA, such correlations are included in the exchange-correlation
kernel, which is static and local in space. Within BSE, they are
included both in quasiparticle energies (through the self-energy
operator) and in the kernels $K^d$ and $K^f$. And these differences
manifest themselves in a very intriguing way in the $\pi-\pi^*$
complex: both theories agree within 0.1~eV in the excitation energy of
components $B_{1u}$ and $E_{1u}$, but they differ by more than 0.5~eV
in the excitation energy of component $B_{2u}$.

\begin{table*}[t]
\caption{Electron affinities in naphthalene, Energies in eV.}
\label{t4.nleneea}
\begin{center}
\begin{tabular}{ccccccc} \hline \hline
 & \hspace{0.cm} LDA
 & \hspace{0.cm} GW$_0$
 & \hspace{0.cm} GW$_f$
 & \hspace{0.cm} DFT(B3LYP) \cite{rienstra-kiracofe01}
 & \hspace{0.cm} OVGF\cite{deleuze03}
 & \hspace{0.cm} Exp.\cite{burrow87} \\ \hline
$b_{1g}$ & 1.38 & -0.09 & -0.60 & & & -0.90 \\
$b_{2g}$ & 2.19 &  0.64 & 0.14 & -0.20 & -1.30 & -0.19 \\
$b_{3u}$ & 0.76 & -1.07 & -1.35 & & & -1.67 \\
$a_{u}$ & -1.14 & -3.03 & -3.44 & & & -3.38 \\
\hline \hline
\end{tabular}
\end{center}
\end{table*}

As observed in Fig. \ref{f4.benzeneqp} and Table \ref{t4.benzeneea}, the inclusion of vertex corrections together
with a TDLDA polarizability is essential for an accurate prediction of
the ionization potential and electron affinity of benzene. In fact,
the GW$_0$ approximation predicts those quantities to be 9.84~eV
and -0.54~eV respectively. Defining a ``HOMO-LUMO gap'' as the
difference between ionization potential and electron affinity, both
GW$_0$ and GW$_f$ approaches agree in the value of the
gap: 10.3~eV. This is somewhat consistent with the observation by del
Sole and collaborators \cite{delsole94}, who conducted a similar
analysis in bulk silicon and found significant shifts in the
valence band maximum and conduction band minimum, but only small change
in the energy gap itself. Due to the cancellation of vertex
contributions, the energy position of the $\pi-\pi^*$ line is not
significantly affected by the choice of self-energy approximation in
the solution of the BSE, as shown in the lower panel of Fig.
\ref{f4.benzenebse}.

\begin{table*}
\caption{Excitation energies of the lowest-energy neutral excitations
  in naphthalene. Spin-singlet states only. Energies in eV.}
\label{t4.nlenebse}
\begin{center}
\begin{tabular}{cccccccc} \hline \hline
 & \hspace{0.cm} TDLDA 
 & \hspace{0.cm} GW$_0$+BSE
 & \hspace{0.cm} GW$_f$+BSE 
 & \hspace{0.cm} TDDFT(B3LYP) \cite{heinze00}
 & \hspace{0.cm}  CASPT2 \cite{rubio93}
 & \hspace{0.cm} Exp. \cite{george68} \\ \hline
$B_{1u}$ & 4.40 & 3.93 & 4.02 & 4.21 & 4.03 & 3.97 \\
$B_{2u}$ & 4.32 & 4.28 & 4.34 & 4.12 & 4.56 & 4.45 \\
$B_{1u}$ & 5.84 & 6.04 & 6.12 & 5.69 & 5.54 & 5.89 \\
$B_{2u}$ & 6.13 & 6.08 & 6.16 & 5.90 & 5.93 & 6.14 \\
\hline \hline
\end{tabular}
\end{center}
\end{table*}

Naphthalene, C$_{10}$H$_8$ is the second smallest oligoacene. For this
molecule, we used the following parameters in the solution of the
Kohn-Sham equation: grid spacing 0.4~a.u., boundary radius 20~a.u.,
and experimental values of bond length. Ionization potentials
associated to all occupied $\pi$ orbitals and some $\sigma$ orbitals
were reported by Dewar and Worley \cite{dewar69},
with the first one being 8.11~eV \cite{dewar69}.
Some of the theoretical calculations of the first ionization potential
are: 7.85~eV (OVGF method \cite{deleuze03}); and 7.59~eV (hybrid
B3LYP/DFT \cite{lin03}). Predictions from the GW$_f$ and GW$_0$
approximations are 8.15~eV and 8.69~eV respectively.
Fig.
\ref{f4.nleneqp} presents a comparison between predictions from the
GW$_f$ and GW$_0$ approximations for all other potentials.
Although both are acceptable, the
GW$_f$ approximation is clearly superior to GW$_0$, giving a maximum
deviation from experiment of no more than 0.3~eV. Vertex corrections
are of the order of 0.6~eV. We should point out
that the magnitude of self-energy corrections for
$\pi$ orbitals (experimental ionization energy ranging from 8 to 13~eV
\cite{dewar69}) is around 2.4~eV, whereas the self-energy corrections
for $\sigma$ orbitals (experimental ionization energy ranging from 13
to 19~eV \cite{dewar69}) is bigger: around 3.6~eV. Since $\sigma$
orbitals are typically deeper in energy than $\pi$, they are expected
to have self-energy corrections larger in magnitude.

There has been some debate in the literature concerning the stability
of the anion C$_{10}$H$_8^-$ \cite{heinis93}. Early electron capture
experiments have observed a stable anion with electron affinity in the
range of 0.15~eV \cite{zlatkis83}. In contrast, electron transmission
spectroscopy measurement have indicated the ion to be unstable, with
negative electron affinity of -0.2~eV \cite{burrow87}. The anion is
predicted by GW$_f$ to be stable, with electron affinity of
0.14~eV. Table \ref{t4.nleneea} presents a comparison between the calculated
electron affinities and the experimental data \cite{burrow87}. The
fact that the anion has such small binding energy
makes any stability analysis, either from theory or experiment, very
difficult. As observed in benzene, there is a number of low-energy
orbitals around the LUMO, most of them highly delocalized. But some of
those orbitals are localized around the molecule, and they
are responsible for the resonant states detected in electron
transmission spectroscopy measurements \cite{burrow87}. For the
resonant states, we find fair agreement in terms of orbital character
and corresponding electron affinity, although there is a discrepancy
of typically 0.2 to 0.3~eV in the last quantity.

The absorption spectrum of naphthalene is found to have a sharp and
well-pronounced line at 6.0~eV. This line is originated from a
$B_{1u}$ transition which is optically active for light polarization
along the line that joins the centers of the two aromatic rings in the
molecule, called ``long molecular axis''. A second, lower absorption
line is located at 4.3~eV and it has $B_{2u}$ character, being highly
active for polarization on the plane of the molecule but perpendicular
to the long molecular axis. TDLDA and BSE give very similar values
for the energy position of both lines, although they differ by 0.5~eV
in the predicted value of the first excited state, a dark $B_{1u}$
singlet excitation.
A comparison between the theoretical calculations and experimental data is
shown in Table \ref{t4.nlenebse}.

In summary, the absorption spectra of benzene and naphthalene are
dominated by a sharp $\pi - \pi^*$ transition, found in both TDLDA and
GW+BSE methodologies. Considering that this transition is composed of
hybridized carbon-$p$ orbitals, it is expected to be very localized in
space. We do not discard the possibility of strong spacial
localization being a major reason for such good agreement between
TDLDA and GW+BSE, despite the conceptual simplicity of the former. The
good agreement between predicted and measured ionization
potentials of benzene and naphthalene is equally remarkable. There has
been a very limited number of cases where such quantities were
calculated within the framework of the GW approximation
\cite{aulbur00,rohlfing00,rohlfing00_1,ismail-beigi03,benedict03,ishii01}, and the present results provide the first benchmarks
for ionization potentials in oligoacenes.

%%%%%%%%%%%%%%%%%%%%%%%%%%%%%%%%%%%%%%%%%%%%%%%%%%%%%%%%%%%%%%%%%%%%%%%%%%
%%%%%%%%%%%%%%%%%%%%%%%%%%%%%%%%%%%%%%%%%%%%%%%%%%%%%%%%%%%%%%%%%%%%%%%%%%

\subsection{Silicon clusters}\label{clusters}

Small semiconductor clusters containing a few tens of atoms and with
various chemical compositions have been mass produced and
characterized in terms of their optical properties and structural
configuration, although the synthesis of small silicon clusters
remains a challenging task \cite{itoh86,mitas01,soloviev01}. From
a theoretical point of view, optical properties of small silicon
clusters are interesting for one aspect: the linear optical spectrum
of bulk silicon is known to be extremely well predicted by a
first-principles approach based on the BSE \cite{onida02,rohlfing00},
whereas TDLDA fails to predict both the optical gap and excitonic
effects \cite{reining02,kim02} \footnote{The inaccurate optical gap
  reflects the gap underestimation inherent to DFT, and it can be
  corrected by an {\it ad hoc} scissors operator. The lack of
  excitonic effects should be attributed to the specific functional
  employed in TDLDA, since other functionals were shown to correctly
  predict the enhancement of the $E_1$ peak relative to $E_2$
  \cite{reining02,kim02}}. On the other hand, optical excitations of
SiH$_4$ and
Si$_2$H$_6$, the two smallest ``clusters'', are correctly described
within TDLDA \cite{vasiliev02}. Based on these two facts, it is
natural to investigate what limits the validity of each theory and
where the crossover size is, beyond which one theory becomes superior
to the other. Such an analysis is challenging because of two major
factors: the limited number of useful experiments, and the complexity
of some numerical implementations of the BSE method.
Recently, a study based on model dielectric screening in silicon
clusters have addressed the issue, with inconclusive results
\cite{benedict03}. Since the present implementation of the BSE
method is particularly efficient for finite systems, we are able to
compare both TDLDA and BSE methods in a wide range of cluster sizes
and with a minimum of {\it ad hoc} assumptions. As we discuss below,
there is unfortunately one additional difficulty in such comparisons:
there is no direct relationship between the optical spectrum of a
nanostructure (quantified by the absorption cross section, for
instance) and the optical spectrum of a macroscopic solid (quantified
by the macroscopic dielectric function).

In general, the surface of synthesized silicon clusters is covered by
passivating particles and, for the smallest ones, they may undergo
significant reconstruction \cite{mitas01}. Our approach is to
construct clusters from fragments of crystalline silicon and adjust the
number of atoms so that the fragment has surface as spherical as
possible. Dangling bonds on the surface are passivated by attaching
hydrogen atoms. We do not consider surface reconstruction. Although
the class of tetrahedral clusters we discuss here may not be the most
stable ones in the size range of 0 to 2 nm, they are expected to be the
most stable ones when the cluster size is very large. Here, we analyze
the clusters SiH$_4$, Si$_5$H$_{12}$, Si$_{10}$H$_{16}$,
Si$_{14}$H$_{20}$, Si$_{29}$H$_{36}$, Si$_{35}$H$_{36}$,
Si$_{47}$H$_{60}$, Si$_{71}$H$_{84}$, Si$_{99}$H$_{100}$, and
Si$_{147}$H$_{100}$. Since
the presence of hydrogen atoms typically requires dense grids, we use
grid spacings ranging from 0.5~a.u. (SiH$_4$) to 0.7~a.u. (Si$_{147}$H$_{100}$) in the solution of the Kohn-Sham equations.

For each cluster, the quasiparticle Hamiltonian is reorthogonalized
in the Kohn-Sham basis. As emphasized in the literature
\cite{rohlfing00,ismail-beigi03}, the self-energy operator is not
diagonal, and Kohn-Sham eigenfunctions are not good approximations for
quasiparticle wave-functions, particularly for the low-energy
unoccupied orbitals in the smallest clusters. There are two main
reasons for this phenomenon: LDA wrongly predicts some of these
unoccupied states to be bound, whereas quasiparticle orbitals are
often unbound and more extended; and the large density of orbitals
within a few eV from the LUMO makes the occurrence of
non-negligible off-diagonal matrix elements of the operator in
Eq. (\ref{e2.hqp}) very frequent.
The Si$_{35}$H$_{36}$ cluster is a typical example. There, overlap
between Kohn-Sham and quasi-particle wave-functions is observed to be
greater than 95\% for most occupied states and for the first few
unoccupied ones, but it reduces to values ranging from 30\% to 90\%
for the higher-energy unoccupied states.

\begin{figure}[hb]
\centering\epsfig{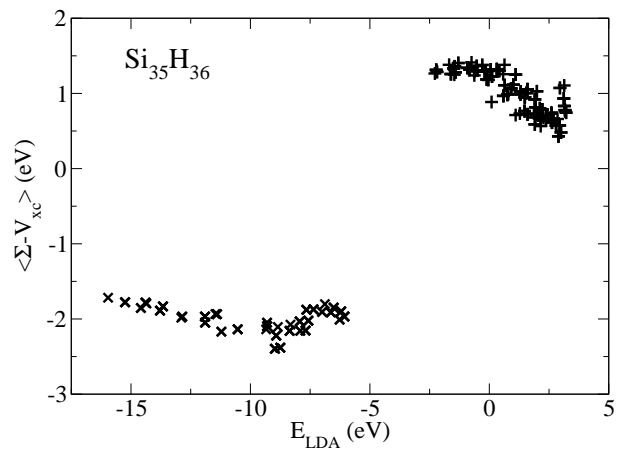}
\caption{Diagonal matrix elements of the operator
  $\Sigma - V_{xc}$, {\it c.f.} Eq. (\ref{e2.hqp}), evaluated in the basis
  of Kohn-Sham eigenfunctions. Matrix elements for occupied
  (unoccupied) orbitals are represented by crosses (``plus''
  signs). Self-energy is evaluated within the GW$_f$ approximation and
  at the extrapolated quasiparticle energy (see text).}
\label{f5.gw35}
\end{figure}

One important aspect involving self-energy corrections is whether they
can be modeled by a ``scissors'' operator or not. Frequently, the
quasiparticle band structure of solids differs from the one predicted
within LDA by a rigid upward shift of conduction bands with respect to
valence bands \cite{hybertsen86,aulbur00}, which justifies the
scissors approximation.  Scissors operators with energy dependence can
also improve band widths \cite{aulbur00}.
But we find this procedure to be inaccurate even for
moderately large clusters. Fig. \ref{f5.gw35} shows diagonal matrix
elements of the self-energy operator as function of the LDA
energy eigenvalue for all occupied orbitals and some unoccupied orbitals of
Si$_{35}$H$_{36}$. For deep occupied orbitals, these matrix elements follow
a quasi-continuous distribution, with smooth curvature, so a
scissors operator could be defined. But the matrix elements
for orbitals in the vicinity of the energy gap do not have a
well-defined dependence with respect to LDA eigenvalues. On the contrary, they
have strong orbital dependence. This fact, together with the existence of large
off-diagonal matrix elements, makes any attempt at constructing a
scissors operator very difficult for these clusters.

\begin{figure}[ht]
\centering\epsfig{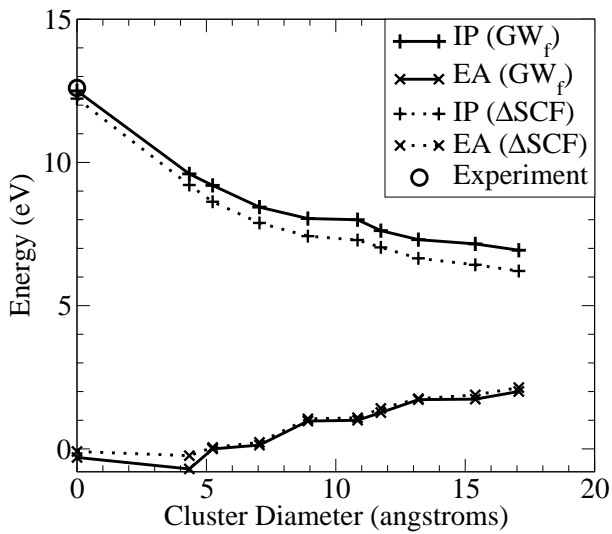}
\caption{First ionization potential and electron
  affinity of passivated silicon clusters, calculated within the
  GW$_f$ approximation (solid lines) and $\Delta$SCF (dotted
  lines). Experimental data from Ref. \cite{itoh86}. $\Delta$SCF
  results include spin-polarization effects.}
\label{f5.gw}
\end{figure}

Fig. \ref{f5.gw} shows the dependence of electron affinity and
first ionization potential with respect to the cluster diameter. To our
knowledge, the ionization potential has not been measured for any of
the studied clusters except for SiH$_4$, for which the experimental
value is 12.6~eV \cite{itoh86}. For this system, the prediction from
GW$_f$ is 12.5~eV. Starting from SiH$_4$, the first ionization
potential decreases monotonically as the cluster size increases. On
the other hand, the electron affinity remains very small, with little
dependence with respect to cluster size. As a result, the electronic
gap (defined as the difference between ionization potential and
electron affinity) decreases continuously for larger and larger
clusters. This is a size effect which has been observed before for
small clusters \cite{rohlfing00}.
In order to understand the physics behind it, we model the cluster as 
an electron sphere with homogeneous density and radius $R$, keeping
the density constant and proportional to $N/R^3$. The energy of the HOMO is found by integrating the density of
states up to the number of electrons $N$, resulting in $E_F =
const. \times N^{2/3} R^{-2}$. The energy difference between this
orbital and the next one is approximately $\Delta E = const. \times
N^{-1/3} R^{-2} = const. \times R^{-3}$, which is inversely
proportional to the volume of the cluster. A more elaborate analysis
predicts a power law $R^{-2}$ instead of $R^{-3}$ \cite{brus83}.

We observe a systematic discrepancy in the first ionization potential
as predicted within GW$_f$ and $\Delta$SCF, the difference being 0.6
to 0.9~eV throughout the range of studied cluster sizes. In contrast,
the electron affinity shows better agreement between these two
theories. Possible explanations
for this behavior are the spurious self-interaction effect, present in
$\Delta$SCF calculations, and the incorrect asymptotic behavior of the
LDA functional \cite{martin,cole82}.

\begin{figure}[ht]
\centering\epsfig{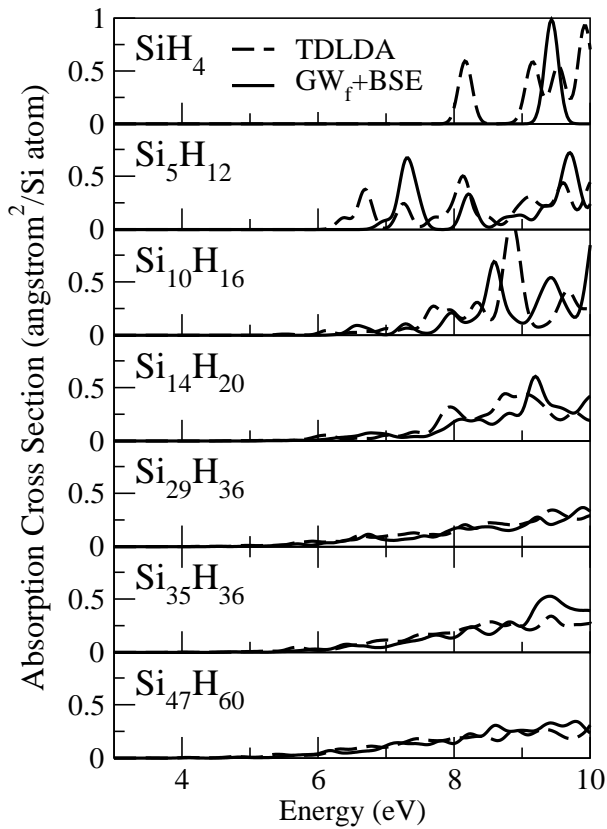}
\caption{Cross section for photoabsorption in
  passivated silicon clusters, calculated within TDLDA (dashed lines)
  and BSE (solid lines). The
  self-energy operator used in BSE is calculated within the GW$_f$
  approximation. The cross section is normalized by the
  number of silicon atoms in each cluster. Absorption lines were
  broadened by a Gaussian convolution with dispersion 0.1~eV.}
\label{f5.abs}
\end{figure}

Fig. \ref{f5.abs} shows the absorption cross section for clusters
SiH$_4$ to Si$_{47}$H$_{60}$. The cross section also has a
characteristic behavior as function of cluster size: for small
clusters, it has a small number of well-defined peaks but, as size
increases, their intensity decreases in the low end of the
energy axis. For large clusters, the absorption cross section shows a
smooth and featureless profile, with onset at a low energy value and
continuous increase towards higher values of energy. Also, TDLDA and
BSE spectra are significantly different for small clusters. In
particular, TDLDA predicts a $s \rightarrow p$ line at around 8.2~eV,
whereas the $s \rightarrow p$ line in BSE is positioned at 9.4~eV. Previous BSE calculations have predicted 9.0~eV \cite{rohlfing98}
and 9.16~eV \cite{rohlfing00} for the same quantity. The
experimental value has been reported to be around 8.8~eV
\cite{itoh86}. For bigger clusters, the differences between TDLDA and
BSE tend to disappear.

As pointed out recently \cite{benedict03}, the behavior of the
absorption cross section for large clusters is dictated by the
exchange kernel $K_{vcv'c'}^x$, which is present in both TDLDA and BSE
formalisms. This term introduces a long-range interaction that
increases in strength as function of cluster size and ultimately
dominates both the exchange-correlation kernel in the TDLDA eigenvalue
equation and the direct+vertex kernel in the BSE. As a result,
oscillator strength is redistributed among the excited states of the
system and the absorption spectrum
peaks at frequencies well above the optical range.
Differences in the detailed treatment of electron-hole
correlations, intrinsic to both TDLDA and BSE, are expected to be less
and less important for increasingly large clusters.

As the cluster size increases, more transitions become allowed and
the absorption cross section is expected to evolve into
a well-pronounced and broad peak at the plasmon-pole
frequency. Indeed, the
formalism in Sections \ref{tdlda} and \ref{bse} can
be applied directly to solids. In solids, Eq.
(\ref{e1.across}) reduces to a function proportional to the imaginary
part of the inverse dielectric function, measurable by electron
energy loss spectroscopy, EELS. Olevano and Reining
\cite{olevano01} have shown that the random-phase approximation
correctly predicts the essential features of the EELS of bulk silicon,
and including corrections at BSE level does not lead to substantial
improvements. As expected, the spectra of the largest clusters
studied, depicted in the lower panels of Fig.
\ref{f5.abs}, show
a smooth profile typical of the low-energy end of the plasmon peak.

The blue shift can be discussed in terms of classical
electrodynamics. We consider a material composed of several identical,
spherical particles in suspension inside an optically inert
background. According to the Mie theory (or {\it effective medium
theory}) \cite{bornwolf,sottile05}, the
(macroscopic) dielectric function of the medium is size-dependent,
and there are three important regimes: particles much smaller than the
wavelength of light; particles of size comparable to the wavelength of
light; and particles much bigger than the wavelength of light. In the
regime of small particles in very low concentration, the imaginary
part of the dielectric function is proportional to the absorption
cross section of each particle according to the expression
\begin{equation}
\epsilon_2 = n \lambda \sigma / 2 \pi \; \; ,
\label{e5.eps2}
\end{equation}
where $\lambda$ is the wavelength of light and $n$ is the
concentration. The two functions then share the same energy
dependence. This is a special case of the Clausius-Mossotti relation
when $n \to 0$. In denser medium, the proportionality above no longer
holds and the energy dependence of $\epsilon_2$ is now given by a
non-linear Clausius-Mossotti relation. In the extreme situation of
highly packed particles, with no interstices, the function
$\epsilon_2$ should resemble the bulk dielectric function. As the
particles condense, any similarity between the energy dependencies of
$\epsilon_2$ and $\sigma$ is lost. The qualitative aspects of this
analysis still hold for more complicated cases, such as non-spherical
and/or inhomogeneous particles. Some predictions of the Mie theory
for the optical properties of nanostructures have been discussed in
recent first-principles calculations: the emergence of a Mie plasmon
in nanoclusters \cite{tsolakidis05}; and the dependence of dielectric
function with respect to concentration in periodic arrangements of
nanowires \cite{zhao04}.

In Fig. \ref{f5.abs}, the parameter is particle size rather than
concentration. From the calculated spectra, one can recover the bulk
limit by increasing the concentration or the size of particles. In the
first case, the Mie theory predicts a red shift of $\epsilon_2$ with
respect to $\sigma$. The second case should certainly involve a
similar phenomenon but it requires a more elaborate analysis, because
of the crossover between particle size and wavelength. The crossover
does not necessarily affect the absorption cross section because light
is not an essential element in absorption measurements (for instance,
one can use electron beams instead of photons as probe). But it should
affect the dielectric function because measurements of dielectric
function necessarily involve interaction with photons.

Although the behavior of the absorption cross section
is well understood, we conclude that spectra of large
clusters are not directly related to the macroscopic dielectric
function.
Whereas
excitonic effects and optical band gaps are easily
recognized in the macroscopic dielectric function, the same does not
hold for the inverse dielectric function. In principle, these two
functions are linked by simple inversion
\cite{hedin69,strinati88,onida02}, but the delicate features of both
functions can be lost if numerical inversion is attempted without
careful control of numerical accuracy. For solids, numerical inversion
is avoided by replacing the BSE for the polarizability operator with a
similar equation that, when solved, provides the macroscopic
dielectric function directly \cite{onida02,hanke78}. For confined
systems, no such prescription is known, and therefore the question of
how to compute an absorption spectrum that, in the bulk limit, reduces
unambiguously to the macroscopic dielectric function remains
unanswered.

The problem of an indeterminate crossover and
ill-defined bulk limit in the absorption spectrum has
two important consequences. The first one is that the identification
of excitonic effects in large silicon clusters becomes a non-trivial
task, even if the cluster is sufficiently large so that the very
concept of ``exciton'' applies. The second is that an optical gap
cannot be easily extracted from the absorption cross section of large
clusters. As seen in Fig. \ref{f5.abs}, the onset of absorption for
the large clusters is not well defined. Often, low-energy
excitations in the cluster have extremely small oscillator strength
and they should not be used to define a gap. Presently, one tentative
definition of the optical gap is via a threshold approach: the optical
gap $E_G$ is such that the integrated cross section up to $E_G$ is a
fixed fraction of the total integrated cross section
\cite{vasiliev02}:

\begin{equation}
\int_0^{E_G} \sigma(E) \dd E = p \int_0^\infty \sigma (E) \dd E \; \; ,
\label{e5.gap}
\end{equation}
where $p$ is a small fraction. This prescription is
motivated by experiment: often, the onset of absorption cross section cannot be
distinguished due to limitations in experimental resolution, and a
practical definition of optical gap becomes important. An obvious
drawback of Eq. (\ref{e5.gap}) is that the gap may be sensitive to the
choice of $p$. In practice, this prescription has been
observed to be robust \cite{vasiliev02}.

\begin{figure}[ht]
\centering\epsfig{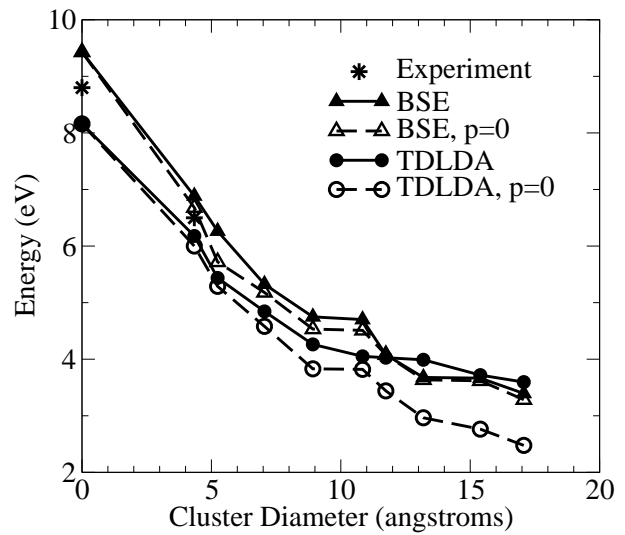}
\caption{Minimum gap (dashed lines and open symbols)
  and optical gap (solid lines and filled symbols) for passivated
  silicon clusters. Circles denote TDLDA gaps. Triangles denote
  GW$_f$+BSE gaps. The optical gap is
  obtained by integrating the oscillator strength up to $p=1 \times
  10^{-4}$. The minimum gap is defined as the lowest
  spin-singlet excitation energy ($p=0$) in the system. Experimental
  data from Refs. \cite{itoh86,delley93}.}
\label{f5.gap}
\end{figure}

In Fig. \ref{f5.gap}, we show the minimum gap and optical gap as
predicted within TDLDA and BSE. The minimum gap is defined as the
energy of the lowest excited state, irrespective of oscillator
strength, whereas the optical gap is defined from Eq.
(\ref{e5.gap}) with $p=2 \times 10^{-4}$. As expected, both gaps have strong
dependence with respect to cluster size. In addition, the BSE gaps are
typically larger than the corresponding TDLDA ones, the difference
slowly decreasing for larger clusters and almost vanishing for
Si$_{147}$H$_{100}$. Fig. \ref{f5.gap} also shows
that the difference between TDLDA minimum and optical gaps has a
slow tendency towards increasing with cluster size, but the same is
not observed in the BSE gaps. The reason for this behavior is that
low-energy excitations obtained within BSE have larger oscillator
strengths than the corresponding ones in TDLDA. Discrepancies between
the TDLDA optical gap in Fig. \ref{f5.gap} and previous work
\cite{vasiliev02} is due to different choices for the value of $p$.

The dependence of the minimum gap with respect to cluster size has also
been discussed in the context of quantum Monte-Carlo (QMC) calculations
\cite{williamson02}. QMC and GW+BSE share a common link: both of
them are many-body theories, and therefore they both include
correlation effects not present in TDLDA. A signature of these effects
can be recognized in the calculated minimum gap: QMC and GW+BSE give
similar values for that quantity, both of them overestimated with respect
to TDLDA ({\it c.f.} Fig. \ref{f5.gap} and
Ref. \cite{williamson02}). For large clusters, there is a tendency for
the QMC gap to fall above the GW+BSE gap. For instance, the GW+BSE
minimum gap for Si$_{147}$H$_{100}$ is found to be 3.3~eV, whereas QMC
predicts a gap of around 4~eV \cite{williamson02}. The corresponding
gap within TDLDA is found to be 2.5~eV.

%%%%%%%%%%%%%%%%%%%%%%%%%%%%%%%%%%%%%%%%%%%%%%%%%%%%%%%%%%%%%%%%%%%%%%%%%%
%%%%%%%%%%%%%%%%%%%%%%%%%%%%%%%%%%%%%%%%%%%%%%%%%%%%%%%%%%%%%%%%%%%%%%%%%%

\subsection{F-center defects in LiCl}\label{licl}

The formalism presented in Section \ref{sigma}
can also be applied to extended systems, once the appropriate boundary
conditions are used in the solution of the Kohn-Sham equations \cite{alemany04}. In
fact, the boundary conditions imposed on electron wave-functions
determine the ones for the polarizability operator $\Pi$, through
Eqs. (\ref{e1.pi_pole}), (\ref{e1.rho}), (\ref{e3.pibsepole}),
and (\ref{e3.bserho}). In periodic systems, optical absorption is
quantified by the absorption coefficient, or by the imaginary part of
the macroscopic dielectric function. Here, we calculate this dielectric
function following an approach due to Hanke \cite{onida02,hanke78}. The
basis of this approach is in defining a truncated Coulomb potential
$\hat{V}$ in the construction of the exchange kernel $K^x$:

\[
K_{vcv'c'}^x \rightarrow \hat{K}_{vcv'c'}^x = K_{vcv'c'}^x - {4
  \pi e^2 \over N_k V_{cell} } { \bra v | \lambda \cdot {\bf v} | c
  \ket \over \varepsilon_c - \varepsilon_v } { \bra c' | \lambda \cdot
  {\bf v} | v'  \ket \over \varepsilon_{c'} - \varepsilon_{v'} } ,
\]
where $\lambda \cdot {\bf v}$ is the velocity operator projected along
some direction $\lambda$. The volume of the periodic cell, $V_{cell}$
and number of k-points, $N_k$, are included as normalization
factors. The kernel $\hat{K}^x$ in the above equation does not have
direct physical meaning, being rather an auxiliary quantity, which
differs from the original $K^x$ by the absence of a long-range Coulomb
interaction. By using this truncated kernel, Hanke has shown
\cite{hanke78} that a polarizability operator can be obtained and,
from that, the macroscopic dielectric function according to

\begin{equation}
Im \left\{ \epsilon \right\} = { 4 \pi^2 e^2 \over N_k V_{cell} }
\sum_s { 1 \over \omega_s^2 } | \int \dd \rr \rho_s (\rr ) \lambda
\cdot {\bf v} |^2 \delta (E - \Omega_s ) \; \; ,
\label{e6.eps}
\end{equation}
where $\rho_s$ and $\omega_s$ can be found by diagonalizing either
 TDLDA-like or BSE-like eigenvalue equations \cite{onida02,kim02,hanke78}.
In addition, lattice periodicity allows us
to use a wave-vector ${\bf q}$ as additional quantum number in the
polarizability $\Pi$. Each choice of ${\bf q}$ leads to a
different eigenvalue equation and the resulting partial
polarizabilities are summed up.

Here, we are interested in analyzing the optical spectrum of a
F center embedded in an otherwise ordered LiCl crystal. F centers are
halogen vacancies found in alkali-halide crystals, and they are named
after the bright, visible coloration they induce. Alkali-halide
crystals are well-known for their strong ionicity and wide energy gap,
which makes them transparent to visible light. Halogen vacancies induce
localized electronic states inside the gap, and electronic transitions
involving such localized states are the source of the bright color in
F-center-rich crystals. The nature of F centers has been studied
experimentally using various techniques \cite{baldini70}.

From a theoretical point of view, F centers are challenging because the
exact location of defect states inside the gap is not easily obtained
without experimental input. Effective-mass models are not suitable due
to the extremely large band gap, high effective masses, and small
dielectric constant. Also, DFT suffers drawbacks due to the
intrinsically underestimated band gap. Within the GW approximation,
not only the band gap is correctly predicted, but defect states around
the F center in LiCl can also be located \cite{surh95}. Here, we
go beyond and
determine optical excitations and the dielectric function within BSE,
and compare the results with the TDLDA prediction and with
experiments.

Defect levels in LiCl are extremely localized, which simplifies
considerably the numerical work. We simulate the defect by using a
cubic supercell with 32 lithium and 32 chlorine atoms. The
lattice parameter is taken from experiment. Starting from the ordered
structure, one chlorine atom is removed and the Kohn-Sham equations are
solved in real space with grid spacing 0.3~a.u. All atoms inside the
periodic cell are allowed to move so as to minimize the total
energy. Only atoms in the first lithium shell showed significant
relaxation, with displacement of 0.05~a.u. and relaxation energy 30~meV. As convergence test, the procedure was repeated in a much bigger
cell, containing 216 atoms of each type (excluding the
vacancy). Movement of atoms in the first shell was around 0.02~a.u. and relaxation energy less than 10~meV, thus showing the degree
of localization of the defect. In the subsequent work, we used the
32Li+32Cl+vacancy cell. Since the presence of a vacancy breaks
translational invariance, we reduced the Brillouin zone to the
$\Gamma$ point only, thus removing complications with Bloch functions
and integrations over the Brillouin zone \cite{alemany04}.

\begin{table}
\caption{Electron band structure of the F center in LiCl, with defect
  levels. All calculated energies are given with respect to the valence band
  maximum as predicted by GW$_f$, in eV. Results obtained from a 2x2x2
  cubic supercell, containing 63 atoms. Defect level $2p$ has a spin
  splitting of 0.4~eV. Results in column ``GW {\it bulk}'' were
  obtained by Hybertsen and Louie \cite{hybertsen86} for bulk LiCl.}
\label{t6.gw}
\begin{center}
\begin{tabular}{lcccc} \hline \hline
 & \hspace{0.cm} LDA
 & \hspace{0.cm} GW$_f$+BSE
 & \hspace{0.cm} GW {\it bulk} \cite{hybertsen86}
 & \hspace{0.cm} Exp. \cite{brown70,baldini70} \\ \hline
 $L_v$ & 0.43 & -0.21 & 0.3 & \\
 $\Gamma_v$ & 0.64 & 0 & 0 & 0 \\
 $1s$ & 5.19 & 6.37 \\
 $2p$ & 8.11 & 10.4-10.8 \\
 $\Gamma_c$ & 6.96 & 9.35 & 9.1 & 9.4 \\
 $L_c$ & 7.06 & 9.63 & 9.7 & \\
 $X_c$ & 8.29 & 10.90-10.7 & \\
\hline \hline
\end{tabular}
\end{center}
\end{table}

Quasiparticle energies for the band edges and two defect levels are
shown in Table \ref{t6.gw}. The choice of supercell allows us to
recognize easily the band edges at points $\Gamma$, $L$, and $X$ by
observing the symmetry properties of the calculated single-particle orbitals.
Self-energy corrections in the neutral F
center require the explicit inclusion of spin degrees of freedom: the
$1s$ state is partially occupied with only one electron. Therefore,
the self-energy operator was evaluated separately for the two possible
spin configurations and the summation over occupied states in Eq.
(\ref{e2.sigmax}) performed over all occupied states for each spin
configuration. A similar procedure was used for the evaluation of the
correlation and vertex parts of the self-energy. Although the
self-energy operator is spin-dependent, most quasiparticle levels
remain spin-degenerate, which is expected since there is
no intrinsic source of spin polarization in the system. Table
\ref{t6.gw} shows that, within DFT-LDA, the $1s$ state is positioned
inside the gap, but the $2p$ triplet state is located within the
conduction band. Inclusion of self-energy corrections keeps the $2p$
triplet above the conduction band maximum (CBM). This phenomenon has
been observed before \cite{surh95}. Although mixed
with extended states, the $2p$ triplet remains fairly localized within
the vacancy. Indeed, 45 \% of its probability distribution is located
within a distance of $L_0$ or less from the vacancy, where $L_0$ is
the Li-Cl interatomic distance. For comparison, 74 \% of the $1s$ probability
distribution is located within the same region.

\begin{figure}[t]
\centering\epsfig{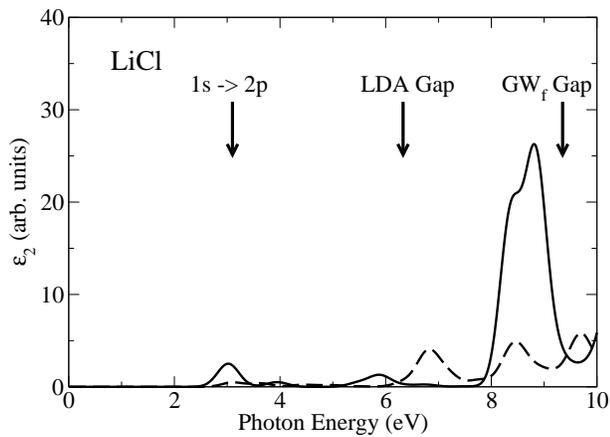}
\caption{Imaginary part of the macroscopic
  dielectric function, $\epsilon_2$ for a F center embedded in
  LiCl. Solid and dashed lines correspond to GW$_f$+BSE and TDLDA
  calculations respectively. The
  self-energy operator used in BSE is calculated within the GW$_f$
  approximation. The position of the $1s \rightarrow 2p$
  transition line and electronic bands gaps within LDA and GW$_f$ are
  shown with arrows. A Gaussian convolution with dispersion
  0.2~eV was used to smear out the absorption lines.}
\label{f6.eps}
\end{figure}

Optical excitations in the system are calculated in both TDLDA and
BSE frameworks. In both cases, spin polarization is included
explicitly in the construction of the electron-hole kernel. Although
not essential at TDLDA level, the inclusion of spin polarization is
important in the BSE because of the existence of the $1s$ partially
occupied orbital. Fig. \ref{f6.eps} shows the imaginary part of the
dielectric function for the cubic supercell containing 63 atoms.
Although this crystal has an electronic band gap of 9.4~eV, the BSE
spectrum is dominated by a wide double peak just below 9~eV. Being
located below the band gap, this peak has the signature of
excitonic effects \cite{rohlfing00}. The limited supercell size
compromises resolution of the excitonic double peak and causes a
redshift of 0.2 to 0.4~eV, although the overall features are consistent
with calculations for a clean LiCl crystal. Below the
excitonic peak, we can see one absorption line corresponding to the
$1s \rightarrow 2p$ transition. In the real material, the strength
of this line depends on temperature and density of defects, and so the
calculated strength is somewhat arbitrary.

The absorption spectrum obtained within TDLDA is characteristic for
the absence of the exciton peak. In fact, it has an onset at the
DFT-LDA band gap, around 6.3~eV, followed by a featureless rise towards higher
energies. The three wide peaks in Fig. \ref{f6.eps}, located at 6.4,
8.3 and 9.7~eV are due to extremely limited resolution in the
supercell. By including more atoms in the supercell, these peaks are
expected to merge into a smooth function. The lack of excitonic
effects in TDLDA has been reported in the literature
\cite{onida02,reining02,benedict03}.

Surprisingly, TDLDA does predict a very accurate value for the $1s
\rightarrow 2p$ transition energy: 3.10~eV, compared to 3.04~eV
(BSE) and the measured value of 3.25~eV to 3.3~eV
\cite{buchenauer68,baldini70,takiyama78,klick65}. We believe this is
related to the strong localization of both $1s$ and $2p$ orbitals. As
discussed before, these orbitals are confined to within 1 or 2 atom
sites from the vacancy. To some extent, the defect behaves as a
confined electronic system, separated from the crystalline
background. For transitions internal to this defect, such as the $1s
\rightarrow 2p$, the LDA exchange-correlation kernel correctly
describes electron-hole interactions. But transitions involving the
conduction and/or valence bands are not so well described because the
LDA kernel lacks the non-locality present in electron-hole
interactions involving extended orbitals. The BSE spectrum also shows
a low feature at around 5.8~eV, which can be assigned to the $1s
\rightarrow L$ line \cite{klick65}.

%%%%%%%%%%%%%%%%%%%%%%%%%%%%%%%%%%%%%%%%%%%%%%%%%%%%%%%%%%%%%%%%%%%%%%%%%%
%%%%%%%%%%%%%%%%%%%%%%%%%%%%%%%%%%%%%%%%%%%%%%%%%%%%%%%%%%%%%%%%%%%%%%%%%%

\section{Conclusion}\label{conclusion}

We discussed an implementation of Green's functions methods developed
in the space of single-particle transitions. Contrary to alternative
approaches, this procedure does not make use of Fourier analysis, and
thus can be applied directly to confined systems, where it is
particularly efficient. As an example, we are able to do a full GW+BSE
calculation of the benzene molecule using a 1.7 GHz IBM Power4 machine
in single-process mode in 280 minutes. This task requires no
more than 600 MB of CPU memory and the only input required is the
geometry of the molecule. Taking advantage of the extreme numerical
efficiency of the current implementation, we are able to perform {\it
  ab initio} GW+BSE calculations of silicon clusters containing more
than one hundred atoms.

Electronic screening is included within the TDLDA, and a corresponding
vertex is included for consistency in the diagrammatic expansion. The
added terms (TDLDA vertex + TDLDA screening) are found to be essential
for accurate predictions of electron affinity and ionization
potentials of benzene and naphthalene. This is in contrast with
previous GW calculations which assume screening at RPA level only but
nevertheless provide accurate ionization potentials for small
molecules \cite{rohlfing00,rohlfing00_1,ismail-beigi03,ishii01}. The
explanation for this apparent contradiction may be that the popularly
used plasmon-pole models carry information about the exact many-body
screening and therefore corrects deficiencies of the RPA.

Besides oligoacenes, the current implementation is used to investigate
the absorption cross section of silicon clusters by solving the
BSE. We conclude that, while the TDLDA and BSE differ markedly for
small clusters, they agree in the broad features of the spectrum for
large clusters. In particular, the dependence of excitation energy as
function of cluster size is similar for both theories. The present
results are consistent with QMC calculations for the minimum gap
of silicon clusters. 
Excitation energies, ionization potentials and electron affinities
calculated within the present method are consistent with experimental
data to within a fraction of eV, comparable to chemical accuracy
\cite{curtiss03}. In LiCl, we
show how the existence of a F center affects the energy-resolved
dielectric function. We also show that TDLDA predicts correctly the
position of the $1s \rightarrow 2p$, despite producing a poor
description of excitonic effects.

\acknowledgments

This work was supported by the National Science Foundation
under DMR-0130395 and DMR-0551195 and by the U.S. Department of Energy
under DE-FG02-89ER45391 and DE-FG02-03ER15491. Calculations were
performed  at the Minnesota Supercomputing Institute (MSI), at the
National Energy Research Scientific Computing Center (NERSC), and at
the Texas Advanced Computing Center (TACC).

%%%%%%%%%%%%%%%%%%%%%%%%%%%%%%%%%%%%%%%%%%%%%%%%%%%%%%%%%%%%%%%%%%
%%%%%%%%%%%%%%%%%%%%%%%%%%%%%%%%%%%%%%%%%%%%%%%%%%%%%%%%%%%%%%%%%%

\appendix

\section{evaluation of energy integral in
Eq. (\ref{e2.sigma0me})}\label{appendix_integral}

The derivation of Eqs. (\ref{e2.sigmax}), (\ref{e2.sigmac}) and
(\ref{e2.vpot}) from Eq. (\ref{e2.sigma0me}) follows from standard
integration over poles \cite{hedin69,hedin95}. For completeness, we
present here the major steps. We start by defining separate exchange
and correlation contributions in Eq. (\ref{e2.sigma0me}):

\begin{equation}
\bra j | \Sigma (E') | j' \ket = \bra j | \Sigma_x | j' \ket + \bra j
| \Sigma_c (E') | j' \ket
\label{e12.sigmaxc}
\end{equation}
with
\begin{eqnarray}
\bra j | \Sigma_x | j' \ket & = \int \dd \rr_1 \dd \rr_2 \varphi_j
(\rr_1 ) i \int {\dd E \over 2 \pi}e^{-iE0^+} \nonumber \\
& \times G (\rr_1,\rr_2 ; E'-E)
\varphi_{j^\prime} (\rr_2) V(\rr_1,\rr_2) \; \; ,
\label{e12.sigmax}
\end{eqnarray}
and
\begin{eqnarray}
& \bra j | \Sigma_c (E')| j' \ket = \int \dd \rr_1 \dd \rr_2 \varphi_j
(\rr_1 ) i \int {\dd E \over 2 \pi}e^{-iE0^+} \nonumber \\
& \times G (\rr_1,\rr_2 ; E'-E)
\varphi_{j^\prime} (\rr_2) \times \nonumber \\
& 
\left[ \int \dd \rr_3 \dd \rr_4
  V(\rr_1,\rr_3) \Pi_0 (\rr_3,\rr_4;E) V(\rr_4,\rr_2) \right] \; \; .
\label{e12.sigmac}
\end{eqnarray}

We assume a quasiparticle approximation for the one-electron Green's
function,

\begin{equation}
G(\rr_1,\rr_2;E) = \sum_n { \varphi_n (\rr_1) \varphi_n (\rr_2) \over
  E - \varepsilon_n + i \eta_n 0^+ } \; \; .
\label{e12.green}
\end{equation}
In the exchange term, Eq. (\ref{e12.sigmax}), the only poles present
are the ones originated from the
Green's function. Therefore, the energy integration is easily replaced
with a summation over occupied states only, resulting in
Eq. (\ref{e2.sigmax}). For the correlation term,
Eq. (\ref{e12.sigmac}), we assume a
polarizability operator given by Eq. (\ref{e1.pi_pole}). Now, both $G$
and $\Pi_0$ contribute with poles below the real energy
axis. Collecting them, one arrives at Eq. (\ref{e2.sigmac}).
%}

\section{static remainder in self-energy}\label{appendix_static}

In most cases, the summation over $n$ in Eqs. (\ref{e2.sigmac}) and
(\ref{e2.sigmaf}) has slow convergence. This behavior
is not unique to the present implementation and it has also been
observed in self-energy calculations when the dielectric matrix is
explicitly computed in reciprocal space
\cite{hybertsen86,aulbur00,reining97}. In that case, converged self-energies
were found to require in excess of 100 unoccupied bands in the
single-particle Green's function. The convergence rate can be
accelerated by truncating this summation at some point and evaluating
the remainder within the COHSEX (Coulomb hole + screened exchange)
approximation \cite{aulbur00}. The
COHSEX approximation is essentially a static limit of Eq.
(\ref{e2.sigma0me}). Assuming that screening is instantaneous, the
polarizability $\Pi_0$ and the potential $W_0$ become proportional to a
$\delta$-function in time and constant in energy. The integral over energy in Eq.
(\ref{e2.sigma0me}) can then be replaced by a sum over poles of the
single-particle Green's function. In our implementation, one can
recover the COHSEX approximation by imposing $\omega_s >> | E -
\varepsilon_n|$ in Eq. (\ref{e2.sigmac}):

\begin{eqnarray}
\left. \bra j | \Sigma_c (E) | j' \ket \right|_{COHSEX} & = 4
\sum_n^{occ} \sum_s { V_{nj}^s V_{nj'}^s  \over \omega_s } \nonumber \\
& - 2 \sum_n
\sum_s { V_{nj}^s V_{nj'}^s  \over \omega_s } \; \; .
\label{e11.cohsex_n}
\end{eqnarray}

The last summation on $n$ is done over all Kohn-Sham eigenstates. We
evaluate it exactly by using a completeness relation. Using
Eqs. (\ref{e1.pi_pole}), (\ref{e1.rho}), (\ref{e1.kx}) and
(\ref{e2.vpot}), we obtain:

\begin{eqnarray}
\left. \bra j | \Sigma_c (E) | j' \ket \right|_{COHSEX} & = 4
\sum_n^{occ} \sum_s { V_{nj}^s V_{nj'}^s  \over \omega_s } \nonumber \\
& + { 1 \over
  2} \int \dd \rr \varphi_j(\rr ) \tilde{W} (\rr ) \varphi_{j'} (\rr )  \; \; ,
\label{e11.cohsex}
\end{eqnarray}
where

\[
\tilde{W} (\rr ) = {1 \over 2} \int \dd \rrp \int \dd \rrpp V(\rr , \rrp
) \Pi_0 (\rrp , \rrpp ; E=0) V(\rrpp , \rr ) \; \; .
\]

Eqs. (\ref{e11.cohsex_n}) and (\ref{e11.cohsex}) can be used to
determine the effect of truncating the sum over $n$ at a value $N >>
n_{occ}$. The remainder then is

\begin{widetext}
\begin{equation}
R_{jj'}^c (N) = {1 \over 2} \int \dd \rr \varphi_j(\rr ) \varphi_{j'}
(\rr ) \int \dd \rrp \int \dd \rrpp V(\rr , \rrp
) \Pi_0 (\rrp , \rrpp ; 0) V(\rrpp , \rr ) + 2 \sum_{n=1}^N
\sum_s { V_{nj}^s V_{nj'}^s  \over \omega_s } \; \; .
\label{e11.wpol0_c}
\end{equation}
\end{widetext}

Although the COHSEX approximation has a level of accuracy lower than
the full
GW method, the convergence behavior is similar in both approximations if
$N$ is chosen sufficiently high. Eq. (\ref{e2.sigmac}) can then
be replaced with

\begin{equation}
\bra j | \Sigma_c (E) | j' \ket =  R_{jj'}^c (N)  + 
2 \sum_{n=1}^N \sum_s { V_{nj}^s V_{nj'}^s
  \over E - \varepsilon_n - \omega_s \eta_n } \; \; .
\label{e11.sigmac_0}
\end{equation}

Along similar lines, the vertex term, Eq. (\ref{e2.sigmaf}) is
rewritten as

\begin{equation}
\bra j | \Sigma_f (E) | j' \ket = \sum_{n=1}^N \sum_s { V_{nj}^s F_{nj'}^s
+  F_{nj}^s V_{nj'}^s \over E - \varepsilon_n - \omega_s \eta_n } +
R_{jj'}^f (N) \; \; ,
\label{e1.sigmaf_0}
\end{equation}
with

\begin{widetext}
\begin{eqnarray}
R_{jj'}^f (N) 
 &
= \sum_{n=1}^N \sum_s { V_{nj}^s F_{nj'}^s
+ F_{nj}^s V_{nj'}^s  \over \omega_s }  + \hspace{8cm} \\
\nonumber
  &
 {1 \over 4}  \int \dd \rr \varphi_j(\rr ) \varphi_{j'}
(\rr ) \int \dd \rr 
\left[ V(\rr , \rrp) \Pi_f (\rrp , \rr ; 0) f(\rr ) +
f(\rr ) \Pi_f (\rr , \rrp ; 0) V(\rrp , \rr ) \right]
\; \; .
\label{e11.wpol0_f}
\end{eqnarray}
\end{widetext}

As an example, we have computed self-energy corrections in benzene
including and not including the static remainder. With a static
remainder, the first ionization
energy decreases by 20~meV when $N$ increases from 256 to
512, resulting in ionization potential of 9.30~eV for the larger
value of $N$. Without static remainder, this ionization energy
increases by 133~meV between the same choices of $N$, and its value for
$N=512$ is 8.27~eV, which is still far from convergence by as much as 1~eV.

In molecules and clusters, high-energy virtual states are expected to
be very delocalized, and therefore sensitive to the choice of
boundary conditions. In spite of that, the use of confined-system
boundary conditions (where all wave-functions are required to vanish
outside some spherical enclosure) is still justified. The reason is
that, as shown in Eq. (\ref{e2.sigmac}), only the overlap between
high-energy and low-energy states is
relevant for self-energy calculations. The detailed features of
high-energy states in the vacuum region, away from the atoms, are not
very important. In addition, the contribution of virtual states in the
summations over $n$ decreases as one goes to higher and higher
states. Nevertheless, the size of the confining region should always be tested
against convergence, so that the shape of virtual states in the
vicinity of atoms is correctly described.

%%%%%%%%%%%%%%%%%%%%%%%%%%%%%%%%%%%%%%%%%%%%%%%%%%%%%%%%%%%%%%%%%%
%%%%%%%%%%%%%%%%%%%%%%%%%%%%%%%%%%%%%%%%%%%%%%%%%%%%%%%%%%%%%%%%%%

%%%%%%%%%%%%%%%%%%%%%%%%%%%%%%%%%%%%%%%%%%%%%%%%%%%%%%%%%%%%%%%%%%%%%%%%%%
%%%%%%%%%%%%%%%%%%%%%%%%%%%%%%%%%%%%%%%%%%%%%%%%%%%%%%%%%%%%%%%%%%%%%%%%%%

\end{document}